\documentclass[11pt]{article}
\usepackage{fullpage}
\usepackage{amssymb}
\usepackage{amstext}
\usepackage{amsmath}
\usepackage{xspace}
\usepackage{graphicx}
\usepackage{graphics}
\usepackage{colordvi}
\usepackage{colordvi}
\usepackage{color}
\newtheorem{lemma}{Lemma}[section]
\newtheorem{theorem}{Theorem}

\newtheorem{claim}{Claim}

\newtheorem{definition}{Definition}
\newenvironment{proof}{\par \smallskip{\bf Proof:}}{\hfill\stopproof}
\def\stopproof{\square}
\def\square{\vbox{\hrule height.2pt\hbox{\vrule width.2pt height5pt \kern5pt
\vrule width.2pt} \hrule height.2pt}}

        {\hspace*{\fill}$\Box$\par\vspace{4mm}}

\newcommand{\opt}{\mbox{\sf OPT}}

\newcommand{\F}{{\mathcal{F}}}

\newcommand{\ignore}[1]{}

\renewcommand{\phi}{\varphi}

\renewcommand{\F}{\ensuremath{\mathbb F}}

\newcommand{\E}[1]{\text{\bf E} \left[#1\right]}
\renewcommand{\d}{{\rm{d}}}

\newcommand{\algoval}{{\text{\sc{LP-Val}}}}
\newcommand{\algogsp}{{\text{\sc{LP-GSP}}}}
\newcommand{\algoposted}{{\text{\sc{LP-Post}}}}

\newcommand{\g}{\varrho}
\newcommand{\tp}{\tilde{p}}

\newcommand{\dist}{\mathcal{D}}

\newcommand{\gsp}{\mbox{\sc Gsp}}
\newcommand{\ogsp}{\mbox{\sc OPT}_{G}}
\newcommand{\oeff}{\mbox{\sc OPT}_{B}}
\newcommand{\app}{\mbox{\sc App}}
\newcommand{\bestkw}{\mbox{\sc BestKW}}

\newcommand{\junk}[1]{}
\date{}
\begin{document}
%
%
%


\title{Selective Call Out and Real Time Bidding}

\author{Tanmoy Chakraborty \thanks{Part of this work was done while
visiting Google Research. Department of Computer and Information Science. University of Pennsylvania,Philadelphia, PA. Email {\tt tanmoy@cis.upenn.edu}}
\and 
Eyal Even-Dar \thanks{Google Research,
76 Ninth Ave, New York, NY. Email: {\tt evendar@google.com}}
\and Sudipto Guha\thanks{Part of this work was done while visiting
Google Research. Department of Computer and Information Science. University of Pennsylvania,Philadelphia, PA. Email {\tt sudipto@cis.upenn.edu}}
\and Yishay Mansour\thanks{Google Israel and The Blavatnik School of Computer Science, Tel-Aviv University, Tel-Aviv, Israel, Email {\tt mansour.yishay@gmail.com}}
\and S. Muthukrishnan\thanks{Google Research,
76 Ninth Ave, New York, NY. Email: {\tt muthu@google.com}}}

\maketitle

\begin{abstract}
Ads on the Internet are increasingly sold via ad exchanges
such as RightMedia, AdECN and Doubleclick Ad Exchange.  These
exchanges allow real-time bidding, that is, each time the publisher
contacts the exchange, the exchange ``calls out'' to solicit bids from
ad networks. This aspect of soliciting bids introduces a novel aspect, 
in contrast to existing literature. 

This suggests developing a joint optimization framework which
optimizes over the allocation and well as solicitation.  We model this
selective call out as an online recurrent Bayesian decision framework
with bandwidth type constraints.  We obtain natural algorithms with
bounded performance guarantees for several natural optimization
criteria. We show that these results hold under different call out
constraint models, and different arrival processes. Interestingly, the
paper shows that under MHR assumptions, the expected revenue of
generalized second price auction with reserve is constant factor of
the expected welfare. Also the analysis herein allow us prove
adaptivity gap type results for the adwords problem.
\end{abstract}



\section{Introduction}
A dominant form of advertising on the Internet involves {\em display}
ads; these are images, videos and other ad forms that are shown on a
web page when viewers navigate to it. Each such showing is called an
{\em impression}. Increasingly, display ads are being sold through
exchanges such as RightMedia, AdECN and DoubleClick Ad Exchange.  On
the arrival of an impression, the exchange solicits bids and runs an
auction on that particular impression. This allows {\em real time
  bidding} where ad networks can determine their bids for each
impression individually in real time (for an example, see
\cite{DCLK}), and more importantly where the creative (advertisement)
can be potentially produced on-the-fly to achieve better
targeting~\cite{rtb}.

This potential targeting comes hand in hand with several challenges.
The Exchange and the networks face a mismatch in infrastructure and
capacities and objectives. From an infrastructure standpoint, the
volume of impressions that come to the exchange is very large
comparison to a smaller ad network limited in servers, bandwidths,
geographic location preferences. This implies a bound on the number of
auctions the network can participate in effectively.  A network would
prefer to be solicited only on impressions which are of interest to
it, and in practice use a descriptive languages to specify features of
impressions (say, only impressions from NY). However this is an
offline feature and runs counter to the attractiveness of real time
bidding. Therefore the exchange has to ``call out'' to the networks
selectively, simultaneously trying to balance the objective of
soliciting as many networks as possible and increasing total value, as
well as not creating congestion or situations where solicitations are
not answered.

\smallskip
This leads to a host of interesting questions in {\em developing a
  joint optimization framework that optimizes over the allocation
  objective as well as the decisions to solicit the bids.}
Specifically, a participant would be solicited for only a
predetermined fraction of impressions. Moreover, these solicitations,
referred to as ``call outs'' henceforth, need to be performed in a
smooth manner and avoid burstiness. The impressions need to be managed
in an online manner, which suggests the use of online algorithms. The
burstiness properties suggests using a queueing model. And the overall
goal is to optimize objectives such as (expected) welfare,
revenue. While each of these issues have been considered in isolation,
the overall challenge is to develop a joint framework, which in turn
raises interesting questions about the interactions between different
parts of the framework.

\smallskip 
The call out framework is formally modeled in Section~\ref{models}.
The online allocation aspect, with a view that the call out constraints
acts as budgets, is reminiscent of the online ad allocation framework
for search ads, or the Adwords problem \cite{MSVV05,BJN,CG08}, and its
stochastic variants \cite{DH09,VVS10}. However the call out framework is 
significantly different, which we discuss below.

The Adwords problem is posed in the deterministic setting where the
expected revenue is treated as a known deterministic reward of
allocating an impression $j$ to an advertiser $i$. The call out
framework has no deterministic analogue; the rationale of the exchange
is that the bids are not known. If the bids were known (or internal)
then we would only call out the winners (assuming multiple slots) of
the auction, which is the path taken by the Adwords problem.  The
call-out framework is similar to Bayesian mechanism
design~\cite{Myerson}. This has some fairly broad conceptual
implications.

First, is the notion of ``adaptivity gap'', where a policy is allowed
to react to realization of the random variables.  The analysis of
adaptivity gap is the central question in the exchange setting.  This
is also relevant in the context of search ads and the adwords setting
where the revenue is achieved on a click which is a random event.  The
adwords model uses the deterministic expectation but it is reasonable
to allow an algorithm to adapt to this event (consider low click
through rates and large bids, such that a payout affects the budget
substantially). To the best of our knowledge, no analysis of
adaptivity gap exists for the adwords problem but such a result will
follow from our analysis. In the call out setting, when optimizing for
welfare or total value, the assignment occurs {\em after} the bids are
obtained, which has considerable gap in comparison to the assignment
that assigns {\em before} obtaining the realizations (reduces to the
expectations).

Second, many objective functions such as generalized second price with
reserve (henceforth GSP-Reserve), for one or multiple slots, have a
very different behavior in the Bayesian and deterministic settings. The
gap between assignment {\em after} and {\em before} the realizations
is more stark in this context -- consider running Myerson's (or
similar) mechanism on the expected bids instead of the
distributions. Note that in GSP-Reserve we announce an uniform reserve
price, {\em before} the bids are solicited as in \cite{Myerson}.
For {\em known} deterministic bids, reserve prices can
be made equal to the bid, and are not useful.  Strong lower bounds
hold for GSP without reserves~\cite{Azar}.

Third, the notion of a comparison class in case of call out
optimization framework requires more care. In the setting of these
large exchanges, a comparison class with full foreknowledge of all
information (in particular, the realization of the bids) is
unrealistic. Moreover, the realizations of the bids depends on the
networks which are called out, and two different strategies that call
out to two different subsets will have completely different
information. Thus to compare two algorithms, it appears that we should
compare their expected outcome -- {\em but each algorithm is allowed to 
be adaptive}. Thus a combination of stochastic and online models are in
order in this setting.

This combination of stochastic and online models relates the call out
framework to the stochastic variants of the Adwords
problem \cite{DH09,VVS10}. But while the similarity implies that
Lagrangian decoupling techniques for separable convex optimization
pioneered by Rockafellar \cite{rockafellar} apply, {\em the different
  possible objectives of the call-out framework are not convex.} In
fact, in the case of optimizing revenue in GSP (with reserve) or in
posted price mechanisms, the objective is not submodular for all prices (as in
welfare maximization). Submodular maximization with linear constraints
has been studied, and while good approximation algorithms exist
\cite{Calinescu,Kulik}, they are inherently offline -- the key aspect
of call out optimization is that the decision has to be made in an
online fashion. The same is true for sequential posted price
mechanisms analyzed in \cite{chawla,sayan} (albeit with more general
matroid setting), the posted prices in the call out setting need to be
announced in parallel (and the eventual allocation is sequential).
Other than formulating the call out framework, significant
contribution of this paper is to demonstrate that relaxations of
natural objective functions can be made separable (as in
\cite{rockafellar}), yet with bounded loss in performance ratio.
Subsequent to the formulation, standard techniques of online
stochastic optimization can be applied.

\subsection{Selective Call Out: The Model}
\label{models}
Let $n$ be the number of ad
 networks $1,2\ldots n$.  We assume that impressions arrive from a
 fixed (unknown) distribution over a
 finite set $U_I$, and that there exists a finite set of bid values
 $U_B$, where $L = \max\{u|\ u \in U_B\}$. In
 the following, ad networks will be indexed by $i\in \{1,2\ldots n\}$,
 impressions by $j\in U_I$ and bid values by $k\in U_B$.
The problem setting involves several steps:

\begin{enumerate}
\item An impression (or keyword) $j$, assumed drawn from a
  distribution $\dist$, comes to the exchange.  There may be {\em
    multiple slots} associated with a single impression, corresponds
  to text ads being blocked together, different locations in the page,
  which are often characterized by different {\em discount rate}. Let
  there be $M$ slots, with discount rates $1\geq \g_1\geq \g_2\ldots
  \geq \g_M\geq 0$.  If a bidder bids $v$, then it is assumed that the
  bid for the $\ell^{th}$ slot is $v \g_\ell$.  The case of $M=1$ is
  common and correspond to a basic pay-per-impression mechanism with
  discount rate $1$. All subsequent discussions apply to this common
  case as well.

\item Given an impression $j$, the bid of ad network $i$ for
  impression $j$ is drawn from a fixed distribution $\mathbf V_{ij}$
  such that the bid is $v$ with probability $p_{ijv}$.  Note that the
  bids of different networks are likely to be correlated based on the
  perceived value of the impression, however, conditioned on $j$, the
  specific dynamics of different bidders can be construed to be
  independent. We assume that the exchange has learned or can predict
  these $p_{ijv}$ given the impression $j$.  This is an assumption
  similar to estimation processes used by search engines to predict
  the value of different advertisements (albeit under static standing
  bids).
\item The exchange decides on the subset $S_j$ of networks to call
  out, subject to the \emph{Call-Out constraints}, which roughly
  bounds the rate at which the exchange can send impressions to
  solicit bid from an ad network. This decision is executed before
  seeing the next impression.
\end{enumerate}

\noindent
To define a specific problem in the above framework, we need to
specify (i) an objective function (ii) a model for call out
constraints, and (iii) the comparison class. The {\bf goal} is to
design a call-out policy that satisfies (ii) and is near optimal in
the objective function (i), when compared to other algorithms in class
(iii). We discuss the instantiations of (i),(ii), and (iii)
in the following.

\smallskip
\noindent
{\bf (i) Objective Functions.} We consider three different objective
functions.  (a) {\em Total Value}: The sum of the maximum bids in
$S_j$ over the arriving impressions $j$.  (b) {\em GSP-Reserve}
(defined above) and (c) {\em Revenue under posted price mechanism
  (take-it-or-leave-it prices)}.  All the quantities are in
expectation. Total value corresponds to the welfare.  The GSP with an
uniform reserve price is a common mechanism used in these
settings. Posted Prices (different networks may get different prices)
are also used in this context. Note that the mechanism is {\em
  parallel} posted price because the prices are posted before any bids
are obtained, and not sequential as in \cite{chawla,sayan}.

\smallskip
\noindent
{\bf (ii) Call Out Constraint Models.} The simplest model for the call
out constraints is a model where one impression arrives at the exchange
at each time step and if the total number of arrivals is $m$, ad
network $i$ can be solicited at most times for some known $\rho_i \leq
1$. We will refer to this as {\em time average model under uniform
  arrival} -- which describes the constraints at the outgoing
and incoming sides of the exchange respectively. Most of the paper
will focus on this model -- {\em primarily because we can show that
  other common models reduce to this variant}. The non-initiated
reader can skip the description of these models and proceed to (iii).

On the incoming side of the exchange, standard practice is to assume
bursty (Poisson) arrivals. We consider this generalization. On the
outgoing side, the simple model allow the possibility that the
call-outs to a network are made on contiguous subset of
impressions. This misses the original goal that the ad network would
receive the impressions at a ``smooth'' rate. A common model used for 
behavior is the {\em token bucket model}
\cite{Tanenbaum}.  A token bucket has two parameters, bucket size
$\sigma$ and token generation rate $\rho$. The tokens represent
sending rights, and the bucket size is the maximum number of tokes we
can store. The tokens are generated at a rate of $\rho$ per unit time,
but the number of tokens never exceeds $\sigma$. In order to send, one
needs to use a token, and if there are no tokens, one can not
send. The output stream of a $(\sigma,\rho)$ token bucket can be
handled by a buffer of size $\sigma$ and a time average rate of $\rho$
-- the buffer is initially full.
Unlimited buffer size corresponds to the time average model.

\smallskip
\noindent
{\bf (iii) Comparison Class.} Given the call out constraints, we define the class of admissible policies.

\begin{definition}
An {\em admissible call out policy} specifies (possibly with
randomization), for each arriving impression $j$, the subset $S_j$ of
ad networks to call out, while satisfying all call out constraints
over the entire sequence of impressions. The policy bases its decision
on the prior information about the bid distributions, and has no
knowledge of the actual bid values. In the case of GSP-Reserve
mechanism, the call out policy also decides the reserve price for each
impression. In the case of a posted price mechanism, the call out
policy also decides the posted/take-it-or-leave-it prices.
\end{definition}

Our comparison class is the set of all admissible call out policies
which know the bid distributions for every impression, but do not know
the actual realization of the bids. The performance of a policy is
measured as the expected (over the bid distributions and the
impression arrivals) objective value obtained per arriving impression,
when impressions are drawn from $\dist$.

\subsection{Our Results, Roadmap and Other related Work}
\label{results-summary}

\vspace*{-0.1in} We provide three algorithms $\algoval$, $\algogsp$,
and $\algoposted$ for the three objectives discussed for the time
average uniform arrival model.  We then prove that any approximation
for any additive objective function translates naturally to an
approximation for the other constraint models.  Recall, $L$ is the
largest possible bid.  {\em A policy is $\alpha$-approximate if it
  achieves at least $\alpha$ times the performance of the optimal
  policy for the corresponding objective.}  The algorithms will have a
natural two-phase approach where we use the $t$ initial impressions as
a sample as exploration and subsequently use/exploit this algorithm
(see Section~\ref{preliminaries} for more discussion).  We show that:

\vspace*{-0.1in}
\begin{theorem}\label{mainthm}
Suppose the optimal policy has expected total value at least
$\delta > 0$. For any $\epsilon>0$, $\algoval$ with a sample of $t
= \tilde{O}(\frac{n^2L}{\delta\epsilon})$ impressions gives a 
$(1-\frac{1}{e} - \epsilon)$-approximate policy.
\end{theorem}

\vspace*{-0.2in}
\begin{theorem}\label{maingsp}
Suppose the optimal GSP-Reserve policy has expected revenue at least
$\delta > 0$. For any $\epsilon>0$, $\algogsp$ with a sample of
$t = \tilde{O}(\frac{n^2L}{\delta\epsilon})$ impressions gives a
$O(1)$-approximate policy, if all
bid distributions satisfy the {\em monotone hazard rate (MHR)}
property. Moreover, the call outs of the policy derived from $\algogsp$ are identical to those of the policy derived from
$\algoval$.
\end{theorem}

\vspace*{-0.1in}
\noindent In particular, we show that when every bidder is solicited
-- GSP-Reserve achieves a revenue that is $O(1)$ factor of optimal
welfare, when all bid distributions satisfy the MHR property. This is
a common distributional assumption in economic theory, and is
satisfied by many distributions \cite{barlow}. This result is in the
same spirit as (but immediately incomparable to) the result in
\cite{sayan}, which relates the optimum sequential posted price
revenue to the optimal welfare under the same assumptions. We are
unaware of such results about GSP-Reserve.

\begin{theorem}\label{postedthm}
Suppose the optimal posted price policy has expected revenue at least
$\delta > 0$. For any $\epsilon>0$, $\algoposted$ with a sample of $t
= \tilde{O}(\frac{n^2L}{\delta\epsilon})$ impressions gives a
$(1-\frac{1}{e} - \epsilon)$-approximate policy. 
\end{theorem}

\noindent 
We do {\bf not} need the MHR assumption unlike the result in
\cite{sayan} for sequential posted prices, since the comparison
classes are different (and the prices are posted in parallel).  We
next show that:

\begin{theorem}\label{tokenthm}
For a given distribution on impressions $\dist$, suppose we have an
$\alpha$-approximate policy for an objective which is
additive given the allocations and the realizations (and is at least
$\delta > 0$) in the time average uniform arrival call out model. Let
$\sigma_i > \sigma\ \forall i$. Then we can convert the policy to a
$(\alpha - \frac{1}{\sigma -1})$-approximate policy in the token
bucket model. The result extends to Poisson arrivals.
\end{theorem}

\smallskip\noindent{\bf Roadmap:} We summarize the results on online
stochastic convex optimization in Section~\ref{preliminaries}. We
subsequently discuss the the total value problem in
Section~\ref{eff}. We discuss the GSP-Reserve problem in
Section~\ref{gspreserve}.  The posted price problem is discussed in
Section~\ref{posted}. The token bucket model and other arrival
assumptions are discussed in Section~\ref{arrival}. To draw a
contrast to the stochastic results, the adversarial order setting is
discussed in Appendix~\ref{adversarial}.

\smallskip\noindent{\bf Other Related Work:}
A combination of stochastic and online components appear in many
different settings \cite{KRT98,Kleinberg05,BIKR07,adaptive} which are not
immediately relevant to the call-out problem. We note that the
bandwidth-like constraints (where the constraint is on a parameter
different than the obtained value, as is the case for call-outs) has
not studied in the bandit setting (see \cite{KSU08,PandeyOlston})
because the horizon is constrained. Finally, bidding and inventory
optimization problems studied in the context of ad
exchanges~\cite{Sergei,Sergei2,MuthuADX}, are not related to the call
out optimization problems.

\section{Preliminaries}
\label{preliminaries}
Consider a maximizing a ``separable'' linear program (LP) ${\cal L}$
defined on $Q$ {\em global} constraints with right hand side $b_i$,
such that the Lagrangian relaxation produced by the transferring these
constraints to the objective function decouples into a collection of
independent non-negative smaller LPs ${\cal L}_j$ over $n'$ variables
and {\em local} constraints. This implies that the objective function
of ${\cal L}$ is a weighted linear combination of ${\cal L}_j$.  The
uniqueness of the optimum solutions for ${\cal L}_j$ implies that
${\cal L}$ reduces to finding the Lagrangian multipliers. The unique
solution is achieved by adding ``small perturbations'', see
Rockafellar~\cite{rockafellar}. However, this approach only provides a
certificate of optimality and a solution, once we are given the
Lagrangians.  The approach does not give us an algorithm to find the
Lagrangian multipliers themselves.

Devanur and Hayes \cite{DH09} showed that if the smaller LPs could be
sampled with the same probability as their contribution to the
objective of ${\cal L}$, and the derivatives of the Lagrangian can be
bounded, then the Lagrangians derived from a small number of samples
(suitably scaled) can be used to solve the overall LP. The weighted
sampling reduces to the prefix of the input if the ${\cal L}_j$s
arrive in random order (see \cite{GM06}). This was
extended to convex programs in \cite{VVS10}. The 
number of sample bound requires several (easy) Lipschitz type properties:

\begin{enumerate}
\item The optimum value of ${\cal L}$ is at least $\delta>0$ and the
optimum solution of ${\cal L}_j$ is at most $R$.
\item For each setting of the Lagrangians, every ${\cal L}_j$ has a unique optimum solution.
\item Reducing $b_i$ by a factor of $1-\epsilon$ does not reduce ${\cal L}$ 
by more than a factor of $(1-\epsilon)$. 
\item ${\cal L}$ does not change by more than a constant times
  $1+\epsilon$ if we alter the value of the optimum Lagrangian multipliers by 
  a factor of $1+\epsilon$. 
\end{enumerate}

\begin{theorem}
\label{meta}
Sample $t=\tilde{O}(\frac{n'QR}{\delta\epsilon})$ of the smaller
linear programs and consider the linear program ${\cal L'}$ which
corresponds to the union of these smaller linear programs and suitably
scaled global constraints. If we use the optimum Lagrangian
multipliers corresponding to the global constraints of ${\cal L'}$ to
solve the decoupled instances of ${\cal L}_j$ as they are available
(in an online fashion) then we produce a $1+\epsilon$ approximation to 
the optimum solution of
${\cal L}$.
\end{theorem}

In the setting of Adwords, the smaller LPs correspond
to the arrival of an impression, and the associated assignment. 
Thus the stochastic framework of the Adwords problem is obviously of
relevance to the call out optimization framework. However the focus
shifts on solving the smaller LPs which encode the call out decision.
A nice outcome of the approach is a simple two phase algorithm;
an {\em Exploration phase} where the samples
are drawn, and an {\em Exploitation phase} where the Lagrangian
multipliers are used.  If $Q,n',R$ are small, then the exploration
phase can be (relatively) short and this yields a natural algorithm.
Thus the goal of the rest of the paper would be to formulate separable convex 
relaxations and achieve the mentioned properties.

\section{The Total Value Problem}
\label{eff}

In this section, we prove Theorem~\ref{mainthm}, and
describe $\algoval$. Let $q_j$ denote the probability that impression
$j$ arrives. We shall add infinitesimal random perturbations to
$p_{ijv}$ which shall not affect the performance of any policy but
ensure $p_{ijv}$ are in general positions, that is, any combination of
them will almost surely create a non-singular matrix.

\noindent{\bf The LP Relaxation: }
\begin{enumerate}
\item 
Let $x_{ij}$ be the (conditional) probability that advertiser $i$ was
  called out on impression $j$. 
\item 
Let $y_{ijv\ell}$ be the probability that advertiser $i$ bid the value $t$ and was assigned the slot $\ell$ (also
  conditioned on $j$). The constraints are named as
  $A(x,y)$ and $B(x,y)$ as shown.
\end{enumerate}

{\small
\begin{eqnarray*}
LP1 = & \max & \sum_j q_j \sum_v \sum_i \sum_{\ell} v \g_{\ell} y_{ijv\ell} \quad \mbox{s.t.} 
\begin{array}{l}
\quad \sum_j q_j x_{ij} \quad \leq \rho_i \\
\left.  \begin{array}{ll}
\sum_i \sum_v y_{ijv\ell} \leq 1 \\
\left.  \begin{array}{ll}
x_{ij} &\leq  1 \\
\sum_{\ell} y_{ijv\ell} & \leq p_{ijv} x_{ij} \\
x_{ij},y_{ijv\ell} &\geq 0
\end{array} \right \} B(x,y)
\end{array} \right \} A(x,y)
\end{array}
\end{eqnarray*}
}

\noindent{\bf Decoupling:} Let $\lambda^*_i$ be the optimum
Lagrangian variable for the constraint $ \sum_j q_j x_{ij} \leq
\rho_i$. $LP1$ then decouples to smaller LPs, $LP2(j,\vec{\lambda^*_i})$ subject to the constraints $A(x,y)$. 
{\small
\begin{eqnarray*}
LP1 = LP1(\vec{\lambda^*_i}) & = & \sum_i \lambda^*_i \rho_i + \sum_j q_j
LP2(j,\vec{\lambda^*_i}) \quad \mbox{where} \quad LP2(j,\vec{\lambda^*_i}) =  \max
\left( \sum_v \sum_i \sum_{\ell} v \g_{\ell} y_{ijv\ell}  -  \sum_i \lambda^*_i x_{ij} \right)
\end{eqnarray*}
}

\noindent\smallskip{\bf Solving $\mathbf{LP2(j,\vec{\lambda^*_i})}$.}
We begin by considering the dual. Let $\tau_{j\ell}$ be the dual of
the constraint $\sum_i \sum_v y_{ijv\ell} \leq 1$. Let $\xi_{ijv}$
correspond to the dual of the constraint $\sum_{\ell} y_{ijv\ell} \leq
p_{ijv} x_{ij}$. Let $\zeta_{ij}$ correspond to the dual of $x_{ij}
\leq 1$.

{\small
\begin{eqnarray*}
DualLP2(j,\vec{\lambda^*_i}) & = & \min \sum_i \zeta_{ij} + \sum_{\ell} \tau_{j\ell} \quad \mbox{s.t.} \quad \begin{array}{ll}
\tau_{j\ell} + \xi_{ijv} & \geq v \g_{\ell} \\
\zeta_{ij} - \sum_v \xi_{ijv}p_{ijv} & \geq - \lambda^*_i \\
\tau_{j\ell},\xi_{ijv},\zeta_{ij} & \geq 0
\end{array}
\end{eqnarray*}
}

\begin{lemma}
\label{structurelemma}
Let $\tau^*_{j\ell}$ be the optimum dual variables for $LP2$. Then
(i) For all $\ell$ there exists $i$ and $v \geq
  \tau^*_{j\ell}/\g_{\ell}$ such that $\xi^*_{ijv} = v\g_\ell
  - \tau^*_{j\ell}$. 
(ii)
$\tau^*_{j\ell}/\g_{\ell}$ is non-increasing in $\ell$.
\end{lemma}
\begin{proof}
For every $\ell$ there must be some $i,v$ such that we have
$\tau^*_{j\ell} + \xi^*_{ijv} = v \g_{\ell}$.  Otherwise we can keep
decreasing $\tau^*_{j\ell}$, keeping all other variables the
same and contradict the optimality of the dual
solution. Now $\xi^*_{ijv} \geq 0$ and the condition on $t$ follows.
The condition corresponds to the set of points
$(v,\xi^*_{ijv})$ in the two dimensional $(x,y)$ plane
being above the lines $\{ y = \g_{\ell} x - \tau^*_{j\ell} \}$.

For the second part, consider $\tau^*_{j\ell}$ and the $i,v$ such that
we have $\tau^*_{j\ell} + \xi^*_{ijv} = v \g_{\ell}$. 
Define $t$ to be the support of $\ell$.
Let $v \geq \tau^*_{j\ell}/\g_{\ell}$ be the largest such support of
$\ell$.  Consider $\tau^*_{j(\ell-1)}$. We have $(\tau^*_{j(\ell-1)} + \xi^*_{ijv})/\g_{\ell-1} \geq v = (\tau^*_{j\ell} + \xi^*_{ijv})/\g_\ell$.
But $\g_{\ell-1} \geq \g_{\ell}$ and thus 
$\tau^*_{j\ell}/\g_{\ell}$ are non-increasing in $\ell$.
Moreover, if $\g_{\ell}=\g_{\ell-1}$ then $\tau^*_{j\ell} = \tau^*_{j(\ell-1)}$ 
(starting from the support of $\ell-1$). 
\end{proof}

\noindent{\bf Decoupling $\mathbf{LP2(j,\vec{\lambda^*_i})}$ itself.}
Consider $LP2(j,\vec{\lambda^*_i})$ with the
Lagrangians $\tau^*_{j\ell}$. The problem decomposes under the constraints $B(x,y)$, to 

{\small
\begin{eqnarray*}
LP2(j,\vec{\lambda^*_i}) & = & \sum_{\ell} \tau^*_{j\ell} + \sum_i  LP3(j,\vec{\lambda^*_i},\vec{\tau^*_{j\ell}},i) \quad \mbox{where} \quad
LP3(j,\vec{\lambda^*_i},\vec{\tau^*_{j\ell}},i) = 
\max \left( \sum_{\ell} \sum_{v} \left[ v \g_{\ell} - \tau^*_{j\ell} \right] y_{ijv\ell}  -  \lambda^*_i x_{ij} \right) 
\end{eqnarray*}
}

\begin{lemma}
\label{optt}
Define $\ell(v) = \arg \max_{\ell'} \left \{ \g_{\ell'} v -
\tau^*_{j\ell'} |\g_{\ell'} v > \tau^*_{j\ell'} \right\}$ and
$\ell(t)=M+1$ if the set is empty.
Set $y^*_{ijv\ell} = p_{ijv}$ if $\ell = \ell(v)$ and $0$ otherwise.
If $\sum_v \sum_{\ell} v \g_\ell y^*_{ijv\ell} \geq \lambda^*_i$ we set $x_{ij}=1$ and $y_{ijv\ell}= y^*_{ijv\ell}$. Otherwise we set $x_{ij} = y_{ijv\ell} = 0$.
\end{lemma}

\begin{proof}
$LP3(j,\vec{\lambda^*_i},\vec{\tau^*_{j\ell}},i)$ is optimized at
  $x_{ij}=1$ or $x_{ij}=0$. This is because if $0 < x_{ij} <1$ and
  $\sum_{\ell} \sum_{v} \left[ v \g_{\ell} - \tau^*_{j\ell} \right]
  y_{ijv\ell} - \lambda^*_i x_{ij} > 0$ then we can multiply all the
  variables by $1/x_{ij}$ and have a better solution. If the
  latter condition is not true then $x_{ij}=0$ is an equivalent
  solution.
If $x_{ij}=1$ the optimal setting for $y_{ijv\ell}$ is
$y^*_{ijv\ell}$.  (Note that $y^*_{ijv\ell}$ is uniquely determined
for a fixed $t$.) Thus the overall optimization follows from comparing
the $x_{ij}=1$ and $x_{ij}=0$ case.
\end{proof}

\smallskip\noindent{\bf Interpretation and the Call Out Algorithm:}
Given $\{\tau^*_{j\ell}\}^M_{\ell=1}$, the distribution $\{p_{ijv}\}$
for $i$, is divided into at most $M+1$ pieces (some of the pieces can
be a single point) given by the upper envelope (the constraint max) of
the lines $\{\g_{\ell} x - \tau^*_{j\ell} \}_{\ell=1}^M$ and the line
$y=0$, in the $x$--$y$ coordinate plane.  Intuitively, seeing the
value $x=t$, if the upper envelope corresponds to the equation
$\g_{\ell} x - \tau^*_{j\ell}$ then we are ``interested'' in the slot
$\ell$.  If the weighted (by $\g_{\ell}$) sum of interests, given by
$\sum_v v \g_\ell y^*_{ijv\ell}$ exceeds $\lambda^*_i$, then it is
beneficial to call out $i$.  We call out based on
this condition and allocate the slots in decreasing order of bids.

\smallskip\noindent{\bf Analysis:}
The LP2 solution satisfies: $LP1=\sum_{i,v} v p_{ijv}\g_{\ell(v)}$ and
$\sum_{i,v:\ell(v)=\ell} p_{ijv} = 1$. For each slot $\ell$, let
$w_i(\ell) = \sum_{v:\ell(v)=\ell} v p_{ijv}$ and $u_i(\ell) =
\sum_{v:\ell(v)=\ell} p_{ijv}$. Order the $i$ in non-increasing order
of $w_i(\ell)/u_i(\ell)$ inside the slot.
If we call out to $i$ and get $t$, then for the sake of analysis we
will consider its contribution to slot $\ell(t)$ only. Moreover, we
stop the contribution to a slot $\ell$ if any any of the $i$ return a
value $t$ with $\ell(t)=\ell$. The best $M$
ordered bids outperform the analyzed contribution in every
scenario. Therefore it suffices to bound the contribution of the
analysis.

\begin{lemma}
\label{easyclaim}
Suppose we are given a set of independent variables $Y_i$ such that
$Pr[Y_i \neq 0]=u_i$ and $E[Y_i] = w_i$. Consider the random variable
$Y$ corresponding to the process which orders the variables $\{Y_i\}$
in non-increasing order of $w_i/u_i$, and stops as soon as the first
non-zero value is seen. Then $E[Y] = \sum_i \prod_{i'<i} (1-u_{i'})
w_i \geq \sum_i w_i \left( 1 - e^{-1}
\right)$.
\end{lemma}
\begin{proof}
Let $F(\{(w_i,u_i)\}) = \sum_i \prod_{i'<i} (1-u_{i'}) w_i$.
Let $\lambda = \sum_i w_i/\sum_i u_i$.
Given the sequence $\{(w_i,u_i)\}$ where $w_i/u_i$ are
non-increasing, if there exists an $i$ such that $w_i/u_i \neq
w_{i+1}/u_{i+1}$, then define a new sequence
$\{(w'_i,u_i)\}$ as follows:

{\small
\[ w'_{i'} = \left \{
\begin{array}{ll}
w_{i'} & \mbox{ if $i' \neq i,i+1$}\\
w_i - \Delta & \mbox{ if $i' = i$}\\
w_{i+1} + \Delta & \mbox{ if $i' = i+1$}
\end{array} \right.
\quad \quad \mbox{where}
\quad \quad \Delta = \frac{\frac{w_i}{u_i} - \frac{w_{i+1}}{u_{i+1}} }{
  \frac1{u_i} +\frac1{u_{i+1}}} \]
}

Note that $\sum_i w_i =\sum_i w'_i$ and $w'_i/u_i$ remains
non-increasing. Now
$ F(\{(w_i,u_i)\}) - F(\{(w'_i,u_i)\}) = \prod_{i'<i} (1-u_{i'})
\Delta - \prod_{i'<i+1} (1-u_{i'}) \Delta =
\prod_{i'<i} (1-u_{i'}) u_i \Delta > 0$.
Thus, we can repeatedly perform the above steps till we get a sequence
such that $w'_i/u_i$ remains the same for all $i$ and $\sum_i w_i =
\sum_i w'_i$. Clearly $w'_i = \lambda u_i$ in this case.
The function $F$ will continue to decrease, and 

{\small
\begin{eqnarray*}
& & F(\{(w_i,u_i)\}) \geq F(\{(w'_i,u_i)\}) = \sum_i \prod_{i'<i}
  (1-u_{i'}) \lambda u_i = \lambda \left( 1 - \prod_i ( 1 - u_i)
  \right) \geq \lambda \left( 1 - e^{-\sum_i u_i} \right)
\end{eqnarray*}
}

But $\frac1{x}(1-e^{-x})$ is decreasing over $[0,1]$ and
the worst case is $x=1$. 
\end{proof}

\noindent In slot $\ell$ (renumbering the advertisers in the order of
$w_i(\ell)/u_i(\ell)$) we get an expected reward of $w_i(\ell)$ if we
reach $i$. But the events are independent in a particular slot. Thus
the expected reward in a slot is bounded by (using independence and
Claim~\ref{easyclaim}) to be $(1-e^{-1})$ times $\sum_i w_i(\ell)$.
We now apply linearity of expectation across the slots -- {\em observe
  that the events across the slots are quite correlated}. The expected
reward is at least $(1- e^{-1})$ times $ \sum_\ell \sum_i w_i(\ell) =
LP1$.  Theorem~\ref{mainthm} follows from Lemmas~\ref{optt} and
\ref{easyclaim} and the application of Theorem~\ref{meta}.

\section{Generalized Second Price with reserve (GSP-Reserve)}
\label{gspreserve}
\newcommand{\spec}{\Psi}

The call outs for this problem would be exactly the
same as the algorithm in Section~\ref{eff}. {\em We will however
  adjust the reserve prices.}  The reserve price will be the same for
all the advertisers being called out on that impression. In fact
either we will run a single slot auction with a reserve price, or
simply GSP for the $M$ slots. The decision will depend on the LP
solution found for this specific impression (and the contributions of
different parts of the LP). 
Recall that the bid distribution $\mathbf V_{ij}$ of advertiser $i$ on impression $j$ is assumed to satisfy the MHR property. We use the following:

\begin{lemma}(Lemma~3.3 in \cite{sayan})
\label{mhrlemma}
For any random variable $V$ following an MHR distribution, let $v^* = \arg \min_v \{ v | v \Pr[ V \geq v] \geq \frac12 \sum_{v' \geq v} v' \Pr[V=v'] \}$. Then $\Pr[V \geq v^*] \geq  e^{-2}$.
\end{lemma}

The next lemma is a restatement of Lemma~\ref{optt} and the subsequent
analysis.

\begin{lemma}
\label{problemma}
Given an impression $j(t)$, and define $v_1(t) = \max_{\ell:\g_1 \neq \g_{\ell} }
(\tau^*_{j1} - \tau^*_{j\ell})/(\g_1 - \g_{\ell})$.  The call out
to a set $S(t) \neq \emptyset$, ensures that $\sum_i \sum_{v \geq
  v_1(t)} p_{ijv} \geq 1$.
\end{lemma}

\begin{definition}
\label{defdef}
Given an impression $j(t)$, and the call out decision to a set $S(t)
\neq \emptyset$ at time $t$, let
$v^*_m(i,t) = \min \{ v |  2 v \Pr[ V_{ij} \geq v] \geq  \sum_{v' \geq v} v' p_{ijv'} \}$ and $\spec(t)= \{ i | i \in S(t) \mbox{ and } v_1(t)  \leq v^*_m(i,t) \}$.
Note that using Lemmas~\ref{mhrlemma}, and \ref{problemma}, we have $|\spec(t)| \leq \lfloor e^2 \rfloor = 7$
since $ i \in \spec(t)$ contributes a probability mass of at least $e^{-2}$.
\end{definition}

\begin{lemma}
\label{problemma3}
Given an impression $j(t)$, and the call out decision to a set $S(t)
\neq \emptyset$ at time $t$, we can set a single threshold $v^*(t)
\geq v_1(t)$ such that if we set a reserve price $v^*(t)$ for a single
slot then the revenue (ignoring the multiplicative discount factor $\g_1$) is at
least $\frac1{4(7e^2 + 1)} \sum_{i \in S(t)} \sum_{v \geq v_1(t)} v
p_{ijv} $.
\end{lemma}

\begin{proof}
Let $\sum_{i \in S(t)} \sum_{v \geq v_1(t)} v p_{ijv} = Z$. We have two cases, (i) $\sum_{i \in \spec(t)} \sum_{v \geq v_1(t)} v p_{ijv} \geq 7e^2Z/(7e^2+1)$ or (ii) otherwise. In case (i), pick the $i \in \spec(t)$ such that $\sum_{v \geq
  v_1(t)} v p_{ijv}$ is maximized, which is at least
$e^2Z/(7e^2+1)$ since $|\spec(t)|\leq 7$.
Let $V_{ij}$ be the random variable that
corresponds to the bid of advertiser $i$ on impression $j$.  Now since
$v^*_m(i,t) \geq v_1(t)$ we have that

{\small
\begin{eqnarray*}
 \sum_{v \geq v^*_m(i,t)} v p_{ijv} & \geq & \Pr \left[V_{ij} \geq v^*_m(i,t) | V_{ij} \geq v_1(t) \right] \sum_{v \geq  v_1(t)} v p_{ijv}
\geq  \Pr \left[V_{ij} \geq v^*_m(i,t) \right] \sum_{v \geq  v_1(t)} v p_{ijv} \leq \frac{1}{e^2} \sum_{v \geq  v_1(t)} v p_{ijv}
\end{eqnarray*}
} 

which is at least $Z/(7e^2+1)$. Now, if we set $v^*(t)=v^*_m(i,t)$
then just from $i$ we have $\sum_{i \in S(t)} \sum_{v \geq v^*(t)}
p_{ijv} \geq \frac12\sum_{v \geq v^*_m(i,t)} v p_{ijv}$ and therefore
in this case the lemma is true.

\smallskip
In case (ii), we have $\sum_{i \in S(t) \setminus \spec(t)} \sum_{v \geq v_1(t)} v p_{ijv} \geq Z/(7e^2+1)$.
But for each $i \in S(t) \setminus \spec(t)$ we have 
$v_1(t) \sum_{v \geq v_1(t)} p_{ijv} \geq \frac12  \sum_{v \geq v_1(t)} v p_{ijv}$
and as a consequence, $v_1(t) \sum_{i \in S(t) \setminus \spec(t)} \sum_{v \geq v_1(t)} p_{ijv}$ is at least $Z/(2(7e^2+1))$.
Consider setting $v^*(t)=v_1(t)$. Let $p = \sum_{i \in S(t) \setminus \spec(t)} \sum_{v \geq v_1(t)}
p_{ijv}$. Since $p \leq 1$ (from definition of $v_1(t)$, see
Lemma~\ref{problemma}) the probability of sale is at least $(1
-\frac1e)p$ which is bounded below by $p/2$. 
The Lemma follows in this case as well.
\end{proof}

\begin{lemma}
\label{problemma4}
Given an impression $j(t)$, and the call out decision to a set $S(t)
\neq \emptyset$ at time $t$, consider (i)
If  $\sum_{i \in S(t)} \sum_v v \g_{\ell(v)} y^*_{ijv\ell(v)} \geq 3 \sum_{i \in S(t)} \sum_{v \geq v_1(t)} v \g_{1} y^*_{ijv1}$ then call-out to $S(t)$ and run regular GSP. 
(ii) Otherwise call-out to $S(t)$ and run a single slot auction with the threshold $v^*(t)$ given by Lemma~\ref{problemma3}.
This algorithm gives a revenue which is $\Omega(1)$ factor of the
LP bound on efficiency which is given by $\sum_v \sum_{i \in S(t)} \sum_v
v \g_{\ell} y^*_{ijv\ell(v)}$.
Note that the call out decisions are based on optimizing the total value/
efficiency of the slots, and thus are feasible.
\end{lemma}
\begin{proof}
Let the non-increasing ordered list of values that are returned for a
time step be $a_1(t)$. 
Suppose we are in case (i). Then the revenue of GSP is at least
$\sum_{r=1}^M \g_r a_{r+1}(t)$. Now since $\g_r$ are decreasing, and
$a_r(t)$ are non-increasing,

{\small
\begin{equation} 
\sum_{r=1}^M \g_r a_{r+1}(t) 
 \geq \sum_{r=1}^M \g_r a_r(t) - \g_1 a_1(t) 
\quad \quad \Longrightarrow \quad
\E{\sum_{r=1}^M \g_r a_{r+1}(t)}
 \geq \E{\sum_{r=1}^M \g_r a_r(t)} - \g_1 \E{a_1(t)}
\label{detailedequation}
\end{equation}
}

\noindent We know from the rounding in Section~\ref{eff} that $\E{\sum_{r=1}^M \g_r a_r(t)}\geq \left(1 - \frac1e \right) \sum_{i \in S(t)} \sum_v v \g_{\ell} y^*_{ijv\ell(v)}$. 

\noindent We observe that $\E{a_1(t)} \leq \sum_{i \in S(t)} \sum_{v
  \geq v_1(t)} v y^*_{ijv1}$. This is easily seen if we write an LP
for the maximum value seen (this LP is for analysis only). Let
$x_{iv}$ be the probability that $i \in S(t)$ is the maximum with
value $v$. Then (we drop the index $j$ for convenience):

{\small
\begin{eqnarray*}
\E{a_1(t)} & \leq & LPMAX=\max \sum_i \sum_v v x_{iv} 
\quad \quad \mbox{s.t.} \quad \quad
\begin{array}{lll}
& & \sum_i \sum_v x_{iv} \leq 1\\
& & x_{iv} \leq p_{iv} \\
& & x_{iv} \geq 0
\end{array}
\end{eqnarray*}
} 

The optimum solution of $LPMAX$ is
$x^*_{iv} = p_{iv}$ for $v > \tau$ and $x^*_{iv} \leq p_{iv}$ for one
$i$ and $v=\tau$. Here $\tau$ is the optimum dual variable for the
constraint $\sum_{i \in S(t)} \sum_v x_{iv} \leq 1$. Note that $\sum_{i \in S(t)} \sum_v
x^*_{iv} = 1$.  

\noindent For $v>v_1(t)$ we have $y^*_{ijv1}=p_{ijv}$ and
$v<v_1(t)$ we have $y^*_{ijv1}=0$. Moreover $\sum_{i\in S(t)}
\sum_{v\geq v_1(t)} y^*_{ijv1} = 1$.
Likewise for $v>\tau$ we have $x^*_{iv}=p_{iv}$ and $v<\tau$ we have
$x^*_{iv}=0$ and $\sum_{i\in S(t)} \sum_{v\geq \tau}
x^*_{iv} = 1$.

\noindent
Suppose that $\tau < v_1(t)$. We arrive at a contradiction because  
$\sum_{i \in S(t)} \sum_{v \geq \tau } x^*_{iv} > $ $ \sum_{i \in S(t)}
\sum_{v \geq v_1(t)} y^*_{ijv1} = 1$ which implies that we are
exceeding the probability mass of $1$ for the maximum.
On the other hand if $\tau > v_1(t)$, then we again have a contradiction that
$ \sum_{i \in S(t)} \sum_{v \geq v_1(t)} y^*_{ijv1} > \sum_{i \in S(t)}
\sum_{v \geq \tau } x^*_{iv} = 1$
which implies $\{y^*_{ijv1}\}$
were not feasible.

\smallskip\noindent
As a consequence, $\tau=v_1(t)$ and for $v>\tau=v_1(t)$ we have 
$x^*_{iv}=y^*_{ijv1}=p_{ijv}=p_{iv}$. For $v<\tau=v_1(t)$ we have 
$x^*_{iv}=y^*_{ijv1}=0$. Therefore
$\E{a_1(t)} \leq LPMAX = \sum_{i \in S(t)} \sum_v v x^*_{iv} = \sum_{i \in S(t)} \sum_v v y^*_{ijv1}$
as claimed.  Applying this claim to Equation~\ref{detailedequation},
and the fact that $\sum_{i \in S(t)} \sum_v v y^*_{ijv1} \leq \frac13
\sum_{i \in S(t)} \sum_v v \g_{\ell} y^*_{ijv\ell(v)}$, we get

{\small
\[ \E{\sum_{r=1}^M \g_r a_{r+1}(t)}
 \geq \left(1 - \frac1e - \frac1{3} \right) \sum_{i \in S(t)} \sum_v v \g_{\ell} y^*_{ijv\ell(v)} \]
}

\noindent Thus in this case the expected revenue is $\Omega(1)$ of the
LP bound on the efficiency.

\smallskip
Suppose we are in case (ii). By Lemma~\ref{problemma3} 
we are guaranteed an expected revenue of $\Omega(1)$ times 
{\small
\[ \g_1 \sum_{i \in S(t)} \sum_{v \geq v_1(t)} v p_{ijv} \geq \sum_{i \in S(t)} \sum_{v \geq v_1(t)} v \g_{1} y^*_{ijv1} \geq \frac13
\sum_{i \in S(t)} \sum_v v \g_{\ell(v)} y^*_{ijv\ell(v)} \]
}
\noindent In this case also the expected revenue is $\Omega(1)$ of the
LP bound; the lemma follows.
\end{proof}

We are ready to prove Theorem~\ref{maingsp}.

\begin{proof}(Of Theorem~\ref{maingsp}).
Let $\app$ be the policy that approximately maximizes the efficiency.
Let $\ogsp $ be the optimum GSP-Reserve policy. 
Let $\oeff$ be the optimum policy which maximizes the total value.

Given a policy $P$ let $\gsp(P)$ denote the expected revenue of the
policy if the charged as GSP, $\bestkw(P)$ denote the expected (weighted)
efficiency. Then for any policy $\gsp(P) \leq \bestkw(P)$.
Let $R(\app)$ be the revenue of the policy in Lemma~\ref{problemma4}.
Therefore, for some absolute constant $\alpha \geq 1$,
\[ \gsp(\ogsp) \leq \bestkw(\ogsp) \leq \bestkw(\oeff) \leq LP1 \leq \alpha \ R(\app)\]
The theorem follows (again appealing to Theorem~\ref{meta}).
\end{proof}

\section{The Posted Price Revenue Problem}
\label{posted}

In this section we prove Theorem~\ref{postedthm}. The flow of ideas would be similar to that of Section~\ref{eff}, however we would require a different LP.
\begin{enumerate}
\item Let $x_{ijv}$ denote the probability that advertiser $i$ was
  called out on impression $j$ with price $t$.
\item Let $y_{ijv\ell}$ be the probability that advertiser $i$
  was offered price $t$ and assigned to slot $\ell$.
  The revenue generated in this event is $\g_{\ell} t$.
\end{enumerate}

Let $\tp_{ijv} = \sum_{t'\geq t} p_{ijv}$ denote the probability that
on impression $j$, advertiser $i$ has a valuation at least $t$. The LP is:

{\small
\begin{eqnarray*}
LP4 = & \max & \sum_j q_j \sum_v \sum_i \sum_{\ell} v \g_{\ell} y_{ijv\ell} 
\quad \mbox{s.t.} \quad 
\begin{array}{lll}
& & \sum_j q_j \sum_v x_{ijv} \leq \rho_i \\
& & \left. \begin{array} {l}
\sum_i \sum_v y_{ijv\ell} \leq 1 \\
\left. \begin{array}{ll}
\sum_v x_{ijv} & \leq 1\\
\sum_{\ell} y_{ijv\ell} & \leq \tp_{ijv} x_{ijv} \\
x_{ijv},y_{ijv\ell} & \geq 0
\end{array} \right\} B(x,y)
\end{array} \right\} A(x,y)
\end{array}
\end{eqnarray*}
}

\smallskip{\bf Decoupling:} Let $\lambda^*_i$ be the optimum dual
variable for the constraint $ \sum_j q_j \sum_v x_{ijv} \leq \rho_i$.
The result of the decoupling is (subject to $A(x,y)$):

{\small
\begin{eqnarray*}
LP4 = LP4(\vec{\lambda^*_i}) =  \sum_i \lambda^*_i \rho_i + \sum_j q_j
LP5(j,\vec{\lambda^*_i})\quad \mbox{where} \quad LP5(j,\vec{\lambda^*_i}) =  \max
\left( \sum_v \sum_i \sum_{\ell} v \g_{\ell} y_{ijv\ell}  -  \sum_i \sum_v \lambda^*_i x_{ijv} \right) 
\end{eqnarray*}
}

\smallskip{\bf Solving $\mathbf{LP5(j,\vec{\lambda^*_i})}$.}
Let $\tau^*_{j\ell}$ be the dual of the constraint $\sum_i \sum_v
y_{ijv\ell} \leq 1$. And again we have a decoupling (subject to $B(x,y)$):

{\small
\begin{eqnarray*}
LP5(j,\vec{\lambda^*_i}) = \sum_{\ell} \tau^*_{j\ell}  +
\sum_i LP6(j,\vec{\lambda^*_i},\vec{\tau^*_{j\ell}},i) \quad \mbox{where} \quad LP6(j,\vec{\lambda^*_i},\vec{\tau^*_{j\ell}},i) & = &
\max \sum_v \sum_{\ell} (t \g_{\ell} - \tau^*_{j\ell} )y_{ijv\ell}  -  \lambda^*_i x_{ijv} 
\end{eqnarray*}
}

\ignore{
\begin{eqnarray*}
LP5(j,\vec{\lambda^*_i}) & = & \sum_{\ell} \tau^*_{j\ell}  + \max
\left( \sum_v \sum_i \sum_{\ell} (t \g_{\ell} - \tau^*_{j\ell} )y_{ijv\ell}  -  \sum_i \lambda^*_i x_{ijv} \right) \\
& = & \sum_{\ell} \tau^*_{j\ell}  +
\sum_i \left( \max \sum_v \sum_{\ell} (t \g_{\ell} - \tau^*_{j\ell} )y_{ijv\ell}  -  \lambda^*_i x_{ijv} \right) \\
& = & \sum_{\ell} \tau^*_{j\ell}  +
\sum_i LP6(j,\vec{\lambda^*_i},\vec{\tau^*_{j\ell}},i) \\
LP6(j,\vec{\lambda^*_i},\vec{\tau^*_{j\ell}},i) & = &
\max \sum_v \sum_{\ell} (t \g_{\ell} - \tau^*_{j\ell} )y_{ijv\ell}  -  \lambda^*_i x_{ijv}  \\
& & \sum_v x_{ijv} \leq 1\\
& & \sum_{\ell} y_{ijv\ell} \leq \tp_{ijv} x_{ijv} \\
& & x_{ijv},y_{ijv\ell} \geq 0
\end{eqnarray*}
}
\noindent The next lemma follows from inspection (assuming $x_{ijv}$
being fixed).

\begin{lemma}
In the optimum  solution to $LP6(j,\vec{\lambda^*_i},\vec{\tau^*_{j\ell}},i)$
we must have $\sum_{\ell} y_{ijv\ell} = \tp_{ijv} x_{ijv}$ and moreover we
should have
{\small
\[ y^*_{ijv\ell} = \left\{ \begin{array}{ll}
\tp_{ijv} x_{ijv} & \mbox{ if $\ell= \arg \max_{\ell'} \{ v \g_{\ell} - \tau^*_{j\ell} |
t \g_{\ell} > \tau^*_{j\ell} \}$}\\
0 & \mbox{otherwise}
\end{array} \right.
\]
}
\end{lemma}

\begin{definition}
\label{slotfunction}
Note that this fixes a {\bf slot function} $f_{ij}(v) = \arg
\max_{\ell'} \{ v \g_{\ell} - \tau^*_{j\ell} | v \g_{\ell} >
\tau^*_{j\ell} \}$ for each $t$. Again note that this corresponds to
the upper envelope of the lines $y = x \g_{\ell} - \tau^*_{j\ell}$ and
$y=0$. Thus $f_{ij}(v)$ can be represented as a piecewise linear
function of $v$. We can view this as ``$i$ is offered price $v$ with the
understanding that if $i$ accepts then $i$ we will attempt to give the
slot $f_{ij}(v)$ to $i$''. Note slot $M+1$ corresponds to $\g_{M+1}=0$.
\end{definition}

Therefore we have: $LP6(j,\vec{\lambda^*_i},\vec{\tau^*_{j\ell}},i) =LP7(j,\vec{\lambda^*_i},\vec{\tau^*_{j\ell}},i)$ where
{\small
\begin{eqnarray*}
LP7(j,\vec{\lambda^*_i},\vec{\tau^*_{j\ell}},i) & = &
\max \sum_v \left[ (t \g_{f_{ij}(v)} - \tau^*_{jf_{ij}(v)} )\tp_{ijv}  -  \lambda^*_i \right] x_{ijv}  \quad \mbox{s.t.} \quad 
\left \{ \begin{array}{lll}
& & \displaystyle \sum_v x_{ijv} \leq 1\\
& & x_{ijv} \geq 0
\end{array} \right.
\end{eqnarray*}
}

\begin{lemma}
In the optimum  solution to $LP7(j,\vec{\lambda^*_i},\vec{\tau^*_{j\ell}},i)$
we have

{\small
\[ x^*_{ijv'} = \left\{ \begin{array}{ll}
1 & \mbox{ if $v' = \arg \max_v (v \g_{f_{ij}(v)} - \tau^*_{jf_{ij}(v)})\tp_{ijv} |
(t \g_{f_{ij}(v)} - \tau^*_{jf_{ij}(v)})\tp_{ijv} > \lambda^*_i \}$ } \\
0 & \mbox{otherwise}
\end{array} \right.
\]
}

Note that we can compute $v'$ easily because $f_{ij}(v)$ can be
represented as a relatively simple function of $v$, based on
Definition~\ref{slotfunction}.
\end{lemma}

Thus we are offering an unique price $v' = \arg \max_v \{ (v
\g_{f_{ij}(v)} - \tau^*_{jf_{ij}(v)})\tp_{ijv} | (v \g_{f_{ij}(v)} -
\tau^*_{jf_{ij}(v)})\tp_{ijv} > \lambda^*_i \}$ to the advertiser $i$
(with the intuitive idea that the advertiser would be considered for
slot $f_{ij}(v')$). We denote $v' = \infty$ if the condition does not
hold for any value of $t$ possible for $i$.  The probability that the
advertiser accepts is $\tp_{ijv'}$ and this is also $ y^*_{ijv'\ell'}$
where $\ell'=f_{ij}(v')$. Therefore given $j$, to advertiser $i$ we
offer a price of $v'(i,j)$ which is a function $i,j$ with a slot
$\ell(i,j)=f_{ij}(v'(i,j))$ in mind.

\smallskip{\bf The Interpretation and the Rounding:} 
We note that for every slot $\ell$ we have
$\sum_{i: \ell = \ell(i,j)} y^*_{ijv'\ell} \leq 1$. And the expected reward is
$\sum_{i: \ell = \ell(i,j)} \g_\ell t'(i,j) y^*_{ijv'\ell}$.

For each slot we order the advertiser in non-increasing order of
$v'(i,j)$, and we can perform the same analysis as
Lemma~\ref{easyclaim}. And for each slot we expect a revenue of
$(1-\frac1e)$ fraction. Since we have one slot in mind for each
advertiser we can sum up across the slots and expect a $(1-\frac1e)$
fraction of the LP revenue. {\bf Note:} that in a given scenario,
after all the bids are accepted, we can make the best allocations and
only increase the revenue in the process. Thus
Theorem~\ref{postedthm} follows.

\section{Handling Bursts: Token Buckets and Poisson Arrivals}
\label{arrival}
\vspace*{-0.1in} We shall now present the proof of Theorem
\ref{tokenthm}.  Recall that the token bucket model starts with a full
buffer -- and thus unlimited buffer size corresponds to the time
average model.

\noindent\smallskip{\bf The Uniform Arrival case:}
We first consider the uniform arrival case, where an impression arrives every
unit time step. Consider any algorithm ${\cal A}$ which makes call outs 
according to the time average model. We define ${\cal A'}$ to behave exactly as ${\cal A}$ except that in the case the token bucket is empty for advertiser $i$, no call out is made to $i$ in ${\cal A'}$.

Let the expected value received by ${\cal A}$ in 
sending impressions to ad network $i$ be $R_i$, so that the total
expected value obtained by our algorithm is $R= \sum_{i=1}^n R_i$.
Let the corresponding values obtained by our algorithm in the token
bucket model be $R_i'$ and $R'$ respectively, and so
$R'=\sum_{i=1}^n R_i'$. We assume that in the
beginning, each token bucket is full of tokens. This is safe to
assume when the process runs for sufficiently long time $T >>
\sigma_i/\rho_i$, since one can simply ignore the impressions
arriving in the first $\sigma_i/\rho_i$ units of time, and lose only
a small fraction of the objective value in expectation.

\begin{lemma}\label{token-mainlem}
$R_i' \geq (1-\frac{1}{\sigma_i-1}) R_i$.
\end{lemma}
\begin{proof}
Let us focus on a single token bucket, that of ad network $i$. We
simply need to calculate the expected fraction of times that an
attempted sending by our algorithm to ad network $i$ succeeds. We
shall show that this fraction is at least
$(1-\frac{1}{\sigma_i-1})$. Given a set of arriving impressions,
every order of their arrival is equally likely, so every impression
is equally likely to have a failed attempt at sending, and so the
value obtained is at least $(1-\frac{1}{\sigma_i-1}) R_i$.

For every arriving impression, the algorithm decides whether to
attempt sending it to ad network $i$. We lower bound the number of
successful sending by the following process: send whenever the
algorithm attempts sending till the token bucket is empty, then
neglect all impressions till the bucket becomes full (dormant
period), then again send whenever the algorithm attempts to do so,
and so on. This can be observed to be a lower bound because for
every impression that the algorithm attempts but fails to send while
the said process succeeds, there must be a unique previous
successful sending by the algorithm during the dormant period of the
process.

At time step $t$, let $X_t$ denote the amount of tokens in the
bucket. For all $t$, $0\leq X_t\leq \sigma_i$, and $X_0 = \sigma_i$.
Let $\tau$ be a stopping time, defined as the first time when $X_t
<1$. It is easy to see that $\E{\tau}<\infty$, since there is a
non-zero probability of a burst of impressions that are sent to ad
network $i$. Let $Z_t = \sigma_i - X_t$, so $Z_0=0$. Note that
$\sigma_i\geq Z_\tau >\sigma_i-1$. Let $Y_t$ be the number of
impressions sent to ad network $i$ before or at time step $t$. Also,
at every time step, let $\alpha_i \leq \rho_i$ denote the
probability that the algorithm sends an arriving impression to ad
network $i$, i.e., $\alpha_i =\sum_j q_j x_{i,j} $. We prove the
following subclaim:

\begin{claim}\label{lem:stoptime}
$\E{Y_\tau} \geq (\sigma_i - 1)^2\alpha_i/\rho_i$.
\end{claim}
\begin{proof}
First, we observe that $Y_t - Y_{t-1}$ is $1$ with probability
$\alpha_i$, and zero otherwise. So $(Y_t -\alpha_i t)$ is a
martingale. Since $\E{\tau}<\infty$, so applying Doob's optional
stopping theorem on the martingale , we get that $\E{Y_\tau -
  \alpha_i \tau}=0$, so $\E{Y_\tau} = \alpha_i \E{\tau}$. We shall now
show  that $\E{\tau}\geq (\sigma_i-1)^2/\rho_i$.

Let $\F_t,\ t\geq 0$ denote
the filtration for the sequence $\{Z_t\}$, that is, $\F_t$ is the
information about all the values $Z_0,Z_1\ldots Z_t$. We now identify
a process $A_t$ such that $Z_t^2  - A_t$ is a martingale, and
$A_0=0$. Clearly, such a process exists (by Doob decomposition), and
$A_{t+1} = \E{Z_{t+1}^2|\F_t} - Z_t^2$. At each time step, $Z_t$
decreases by $\rho_i$ (lower bounded by zero), and increases by $1$
(upper bounded by $\sigma_i$) with probability
$\alpha_i$. Thus, when $\sigma_i - 1 \geq Z_t \geq \rho_i$,

{\small
\begin{equation*}
\begin{aligned}
\E{Z_{t+1}^2|\F_t} &= (1-\alpha_i) (Z_t - \rho_i)^2 +
\alpha_i (Z_t + 1-\rho_i)^2\\
&=(1-\alpha_i) (Z_t^2 - 2 Z_t \rho_i + \rho_i^2) +\\ 
& \ \ \ \ \ \ \alpha_i (Z_t^2 + 2Z_t (1-\rho_i) + (1-\rho_i)^2)\\
&= Z_t^2 + 2(\alpha_i - \rho_i) Z_t + \alpha_i(1-\rho_i)^2 +
(1-\alpha_i)\rho_i^2\\
&\leq Z_t^2 + \rho_i^2 + \alpha_i (1-2\rho_i)
\leq Z_t^2 + \rho_i
\end{aligned}
\end{equation*}
}

The inequalities follow from the facts that $\alpha_i\leq \rho_i\leq
1$ and $Z_t\geq 0$. Moreover, when $\rho_i\geq Z_t\geq 0$, $Z_{t+1}$
is zero with probability $(1-\alpha_i)$, and at most $1-\rho_i + Z_t \leq
1$ with probability $\alpha_i$. Thus $\E{Z_{t+1}^2|\F_t} \leq
\alpha_i\leq \rho_i$. This implies that $A_{t+1} - A{t}<\rho_i$ and so
$A_t <\rho_t$, for all $t<\tau$.

Now we apply Doob's optional stopping theorem to obtain that
$\E{Z_{\tau}^2 - A_{\tau}} = 0$. Since $Z_{\tau}>\sigma_i -1$, we have
$
(\sigma_i -1)^2  - \rho_i \E{\tau} \leq 0
$. This proves the claim.
\end{proof}

The process will succeed in all attempts to send
impressions to ad network $i$ till time $\tau$. After this, the
process is dormant and neglects all impressions for the next
$Z_{\tau}/\rho_i$ steps, so that the bucket becomes full, and then
the same argument can be repeated. During the dormant period, the
expected number of attempted sending by the original algorithm is at
most $Z_{\tau} \alpha_i/\rho_i$. So the fraction of sending attempts
that are successful is at least

{\small
\[
\frac{\frac{(\sigma_i -1)^2 \alpha_i}{\rho_i}}{\frac{(\sigma_i - 1)^2
    \alpha_i}{\rho_i} + \frac{\sigma_i \alpha_i}{\rho_i}} > 1 -
\frac{1}{\sigma_i-1} \  .
\]
}

This completes the proof of Lemma \ref{token-mainlem}.
\end{proof}

Lemma \ref{token-mainlem} implies that if $\sigma_i > \sigma\ \forall
i$, then $R'\geq (1-\frac{1}{\sigma -1})R$. 

\smallskip\noindent{\bf Poisson Arrival Process.}
%
Now we consider the token bucket model under the assumption that 
the time difference between the arrival of two consecutive impressions
is drawn from an exponential distribution (with mean $1$, for
normalization). This implies that the arrival process is a unit rate
{\em Poisson process}. The tokens fill up
{\em continuously}, at a rate of $\rho_i$ tokens per unit time. 
Extending our analysis from discrete to continuous martingales, in proof of 
Lemma~\ref{token-mainlem}, we get,

\begin{lemma}\label{token-Poissonlem}
$R_i' \geq (1-\frac{1}{\sigma_i-1}) R_i$ in the Poisson arrival process as well.
\end{lemma}
\begin{proof}
$X_t$, $Y_t$ and $Z_t = \sigma_i - X_t$ are now continuous
processes. The Doob optional stopping theorem holds for continuous
martingales, and we can find a process $A_t$ such that $Z_t^2 - A_t$
is a continuous martingale. Let $\tau$ be again the smallest $t$
such that $X_t<1$. It again boils down to showing that $\E{\tau}\geq
(\sigma_i - 1)^2/\rho_i$. To show this, we need to show that
$A_t<\rho_i$ for all $t<\tau$. Between time $t$ and $t+\d t$, where
$\d t$ is an infinitesimal increment, there is a probability $\d t$
that an impression arrives (since the arrival time is a Poisson
process), and thus probability $\alpha_i \d t$ that the algorithm
attempts to send an impression to ad network $i$, while $\rho_i\d t$
tokens are generated. Thus, when $0<Z_t<\sigma_i -1$, we have
\text{(neglecting higher powers of \d t)}:

{\small
\begin{equation*}
\begin{aligned}
\E{Z_{t+\d t}^2|\F_t} &= (1-\alpha_i\d t) (Z_t - \rho_i\d t)^2 +
\alpha_i \d t (Z_t + 1 -\rho_i\d t)^2&\\
&= Z_t^2 + \alpha_i\d t -2Z_t (\rho_i - \alpha_i) \d t
\leq Z_t^2 + \rho_i \d t\\
\end{aligned}
\end{equation*}
}

So we infer that $A_{t+\d t} - A_t \leq \rho_i \d t$, and so
$A_t<\rho_i t\ \forall t$. Finally, applying Doob's optional stopping
theorem for continuous martingales, we get that $\E{Z_{\tau}^2 -
  A_{\tau}} = 0$, which implies that $(\sigma_i - 1)^2 -
\rho_i\E{\tau} <0$.
\end{proof}

This completes the proof of Theorem \ref{tokenthm}.

\section{Experimental Study}
In this section we experimentally
explore aspects of our model and performance of algorithms,
focusing
solely on {\em sales} as our measure of
performance in our simulations, and sticking to the basic mechanism
where ad networks report their bids. In each online query, there is a single impression with discount factor $1$, and a minimum price set by the publisher. The impression is considered to be sold if at least one of the ad networks that are called out returns a bid higher than this minimum price. The objective is to maximize the number of impressions sold. Sales can be viewed as a special case of the efficiency problem, where all bids are $0$ or $1$ (they are willing to buy the item at a minimum price quoted by the publisher, or not), and so the corresponding algorithm is applicable. The following questions are important.

\begin{enumerate}
\item
How much does estimating the bid distributions help?
We assumed that the Exchange will estimate the {\em survival probability} $p_{ijk}$'s for ad network $i$,
impression $j$ and bid $k$ (that is, probability that the bid is at least $k$), via machine learning and data mining techniques.
Can we require less out of these techniques?
Motivated by sponsored search systems, it
is tempting to only estimate the expected bid for each $i,j$, rather than the
entire distribution of $p_{ijk}$. Further, for the sales metric, it suffices to
only estimate the probability of a bid above $k$ for $i,j$, while for the value metric,
we use the entire distribution of $p_{ijk}$.
We compare performance of algorithms that use different amounts of information
about $p_{ijk}$'s.

\item {\em How does the error in estimation affect the performance of algorithms?}
Methods that estimate $p_{ijk}$ will have errors. We study the influence
of such errors on our algorithms.

\item {\em What is the benefit of the optimization over simple, natural
    schemes?}
We evaluate our LP-based solution and compare it to other simpler schemes.
\end{enumerate}

Before proceeding to the simulation setup, we list the algorithms we
consider. The linear programming based algorithm uses probability
thresholds to decide on call outs. However alternative (and simpler)
implementations typically would choose subsets based on some criterion.
Thus we have a natural partition of set and threshold based algorithms.~\\

{\bf Set Based Algorithms:} All our candidate algorithms sort the ad
networks according to some criterion, which may depend on the arriving
impression as well as history of the process. The algorithms then attempt
to call out to the top ad networks in this ordering, which succeeds if
those ad networks have bandwidth remaining. It remains to specify how many
ad networks are picked for a particular impression. Set based algorithms
are simply those that pick the top $k$ ad networks for every impression.
The first two algorithms are baseline algorithms that are motivated by
scheduling algorithms, but they ignore the probability estimations. The
next two algorithms use the probability estimation to guide the call outs
(and are based on different but widely occurring ways of characterizing
high probability subsets).
\begin{enumerate}
\item {\bf Random:} chooses a random ordering.

\item {\bf MaxRemBand:} orders by (decreasing) remaining bandwidth.

\item {\bf MaxProb:} orders by (decreasing) survival probability.

\item {\bf MaxExp:} orders by (decreasing) expectation of a bid.
While expected bid value can be clearly shown
  to be inadequate on specially designed examples, we evaluate the
  value of this information on our more generally generated data sets.
\end{enumerate}

\smallskip \noindent {\bf Threshold Algorithms: } The above algorithms all
have a parameter $k$, the size of the set of ad networks to which call out
is attempted for each impression (subject to bandwidth availability).
Each of the algorithms in this class induce an ordering among ad networks
and choose a prefix of this ordering such that the survival probabilities
$\sum_{i} p_i x_i$
of the chosen ad networks add up to a {\em threshold}.
If the sum exceeds the threshold, then the
algorithm probabilistically decides whether to attempt call out to the
last ad network in the chosen ordering. The probability of
this event is set such that the sum of survival probabilities times the
probability of getting called out sums up to {\em exactly} the threshold.
Note that our LP-based algorithm is close to a threshold algorithm with
the threshold set at $1$, except that the Lagrangians $z_i^*$ that were
learned also acts as a {\em cut-off}.  We found this algorithm to be
conservative in spending the call outs and decided to set this to a
variable threshold.  The class of threshold algorithms is justified by the
Lemma \ref{lem-basic-sales}. Below are all the threshold algorithms that we consider (again, we only
  make the call outs if the actual bandwidth is available).
\ignore{
Thus, if $\sum_{i=1}^n p_i x_i$ adds up to a fixed threshold, then
there is a constant probability of selling the ad network, and it
makes sense to spare the bandwidths of other ad networks for future
impressions. Below are the threshold algorithms that we consider. We
analyze their performance as the threshold is varied.
}

\begin{enumerate}
\item {\bf Th-Random:} chooses a random ordering.
\item {\bf Th-MaxRemBand:} orders on maximum remaining bandwidth.
\item {\bf Th-Prob:} orders on (decreasing) survival probability.
\item {\bf Th-LP:} This is the LP based algorithm
  discussed above.
\end{enumerate}

\begin{figure*}[ht]
\begin{tabular}{cc}
\includegraphics[scale=0.2]{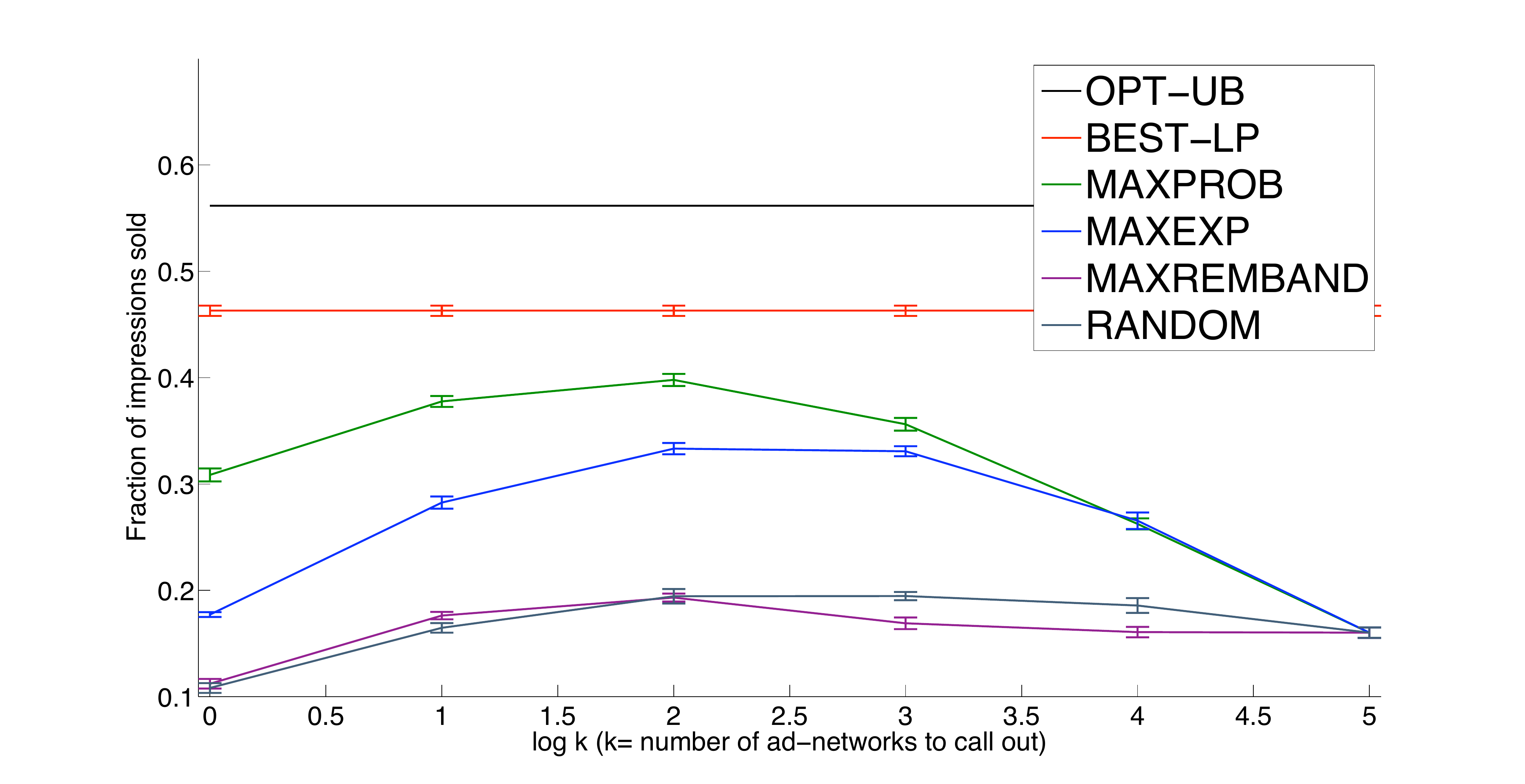}
&
\includegraphics[scale=0.2]{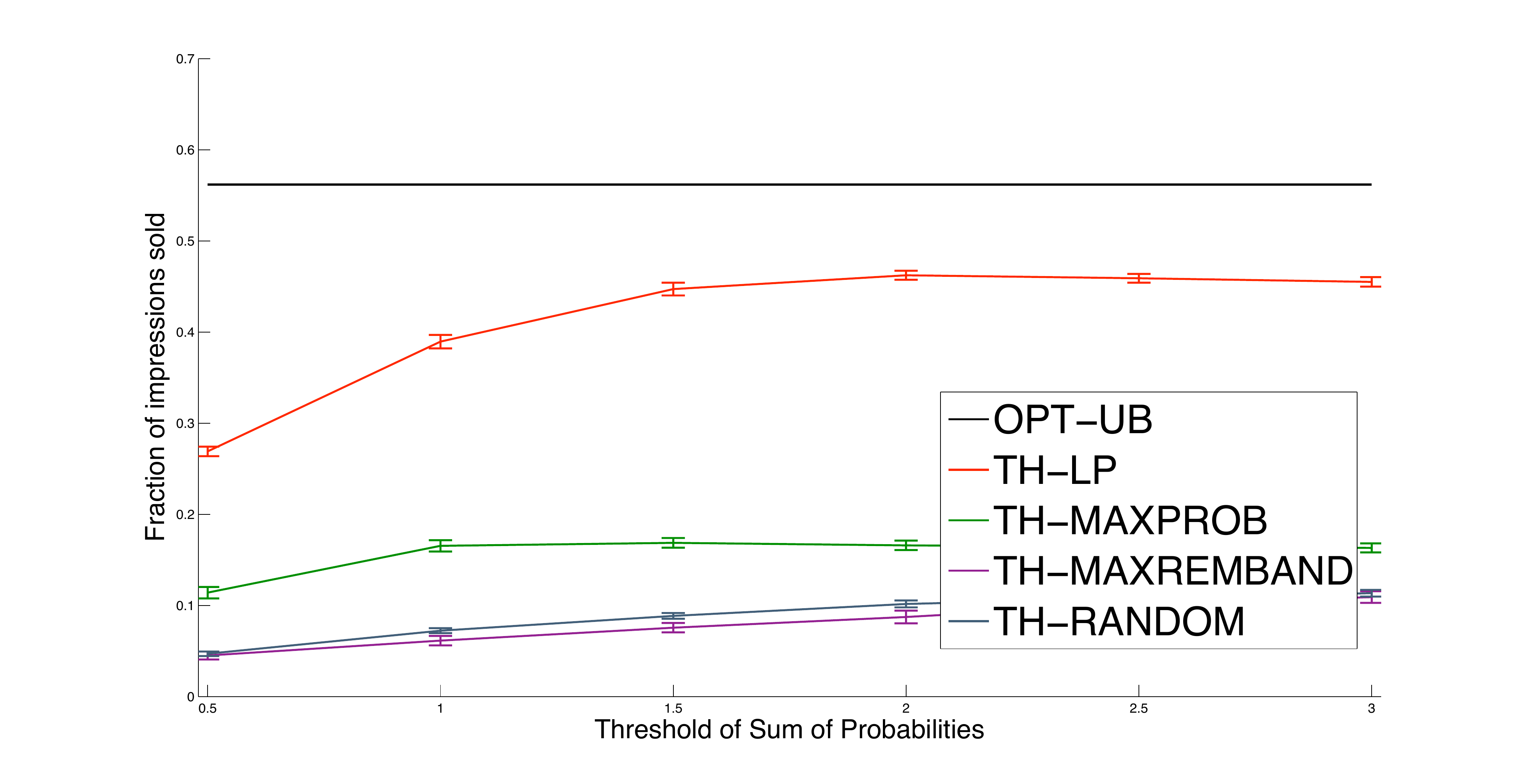}
\end{tabular}
\caption{Set based and Threshold algorithms for Gaussian bid distributions}\label{G0000}
\end{figure*}

\begin{figure*}
\begin{tabular}{cc}
\includegraphics[scale=0.2]{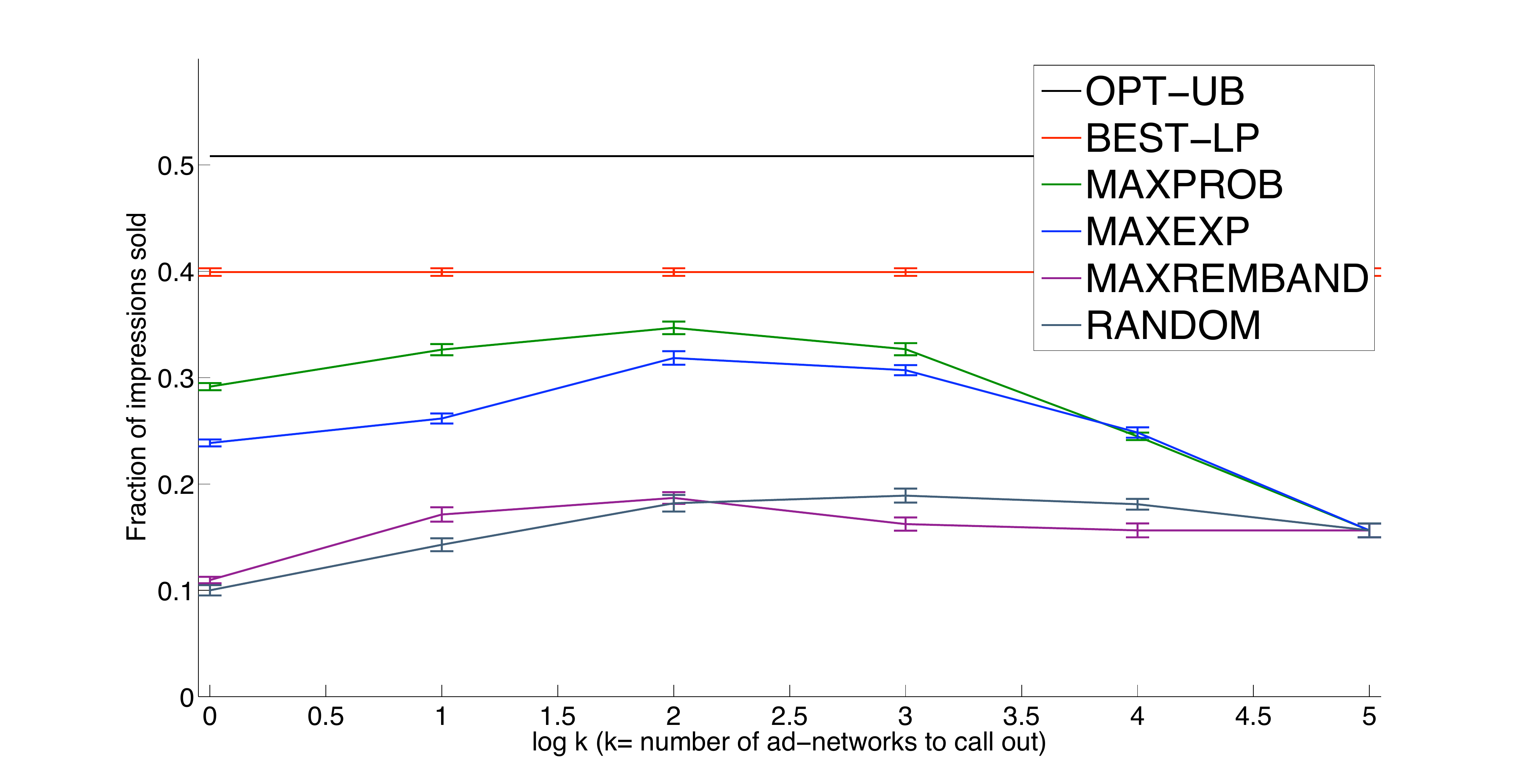}
&
\includegraphics[scale=0.2]{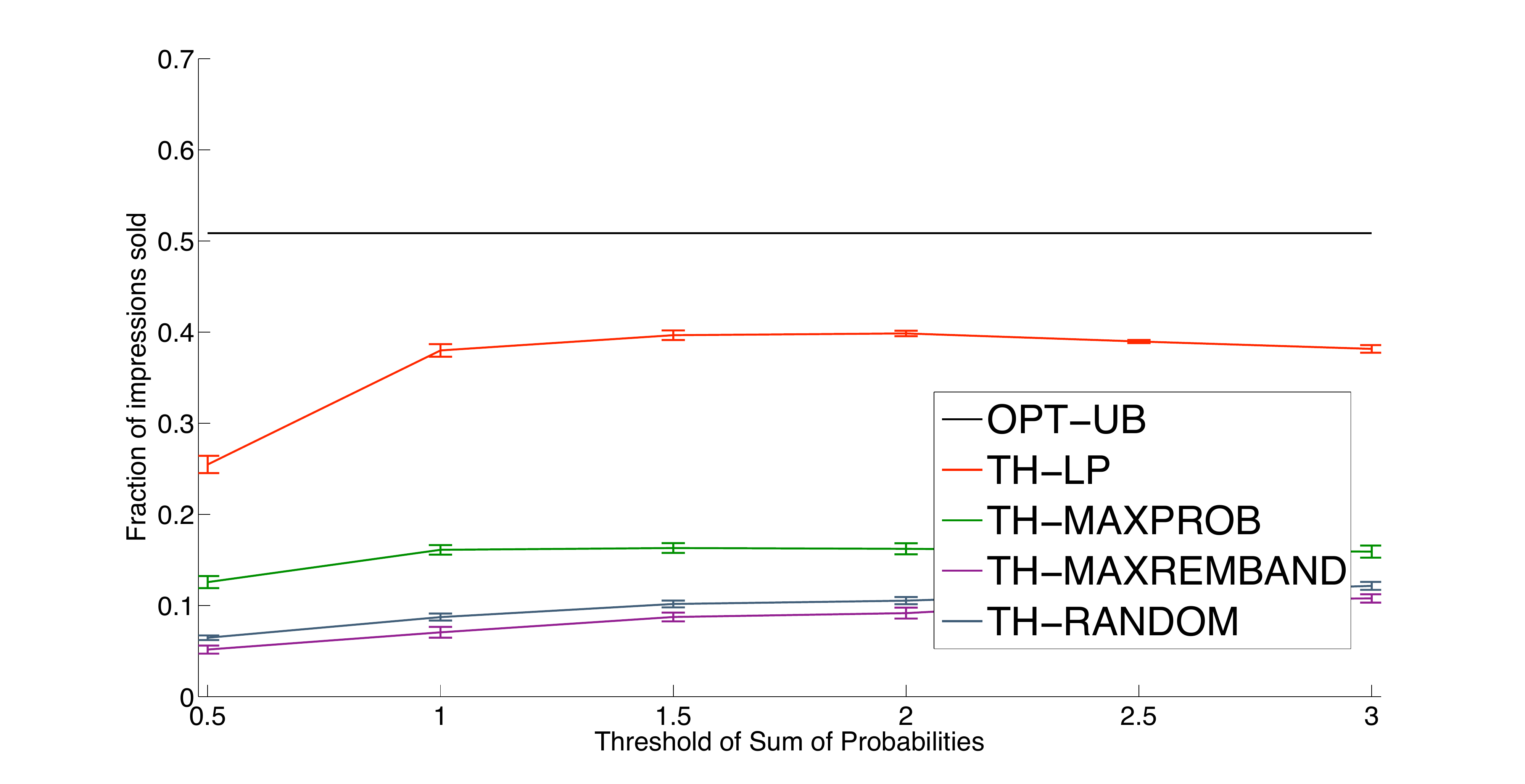}
\end{tabular}
\caption{Set-based and Threshold algorithms for Pareto bid distributions}\label{P0000}
\end{figure*}

\subsection{The Simulation Setup}

We simulate the above algorithms using synthetic data generated from
specific natural distributions, and vary parameters in the data
generation such as impression arrival rate and range of minimum prices
of impressions.

{\bf Implementation:} There are $32$ ad networks, and the bandwidth
of each ad network is implemented as a {\em token bucket} or a
buffer. Tokens get generated in a bucket at a uniform rate, which
reflects the bandwidth, and an attempt to call out to an ad network
succeeds if and only if there is at least one token, which is
consumed by the call out, in the bucket corresponding to the ad
network. Moreover, a token bucket has a limit on the number of
tokens it can store, and if the bucket is full, tokens generated at
that time are lost. This reflects the burst-size allowed in the
communication. Unless specified otherwise we use a bucket size of
$5$ for most of our simulation results (see
section~\ref{sensitivity}). The ad networks have (continuous) token
generation rates chosen uniformly at random between $5$ and $50$ per
unit time. Impressions arrive according to a Poisson clock with a
fixed rate. This rate is the expected time lapse between two
consecutive impressions. Again, unless specified otherwise, we use a
rate of $0.003$ for the Poisson clock. The average token generation
rate per impression, $\rho_i$, varies from $0.015$ to $0.15$ when
the rate of the Poisson clock is $0.003$.

{\bf Bid distribution:}
All bids are drawn from bounded distributions, which put all its
probability mass in the range $[0,R]$ for a fixed positive value
$R$.  Gaussian is a very commonly seen distribution, with exponential
decay, while Pareto is a heavy-tailed distribution which has
polynomial decay ({\em power law}), and is also often observed in
online ad scenarios. The means of these Gaussian or Pareto
distributions are chosen uniformly at random from the range
$[0,0.5R]$. The Gaussian distributions are then given a standard
deviation uniformly between $0$ and $0.5$ times the mean, while the
degree of the polynomial pdf of Pareto distributions is uniformly
chosen between $2$ and $5$. Subsequent to their choice, all
distributions were truncated to $[0,R]$ -- note that the truncation
affects the expectation and survival probabilities.

{\bf Verticals and Minimum Prices:}
Different types of impressions should have different value to an ad
network, so we have $10$ different types of impressions, which we
refer to as {\em verticals}. Each arriving impression is assigned a
vertical uniformly at random. The bid distribution of each ad network
for an impressions depends on its vertical only. However, the survival
probabilities can still vary for impressions from the same vertical,
because their minimum prices are independently chosen. Thus,
impressions are iid drawn from a fixed distribution. $2000$
impressions arrive in each simulation, while our LP-based algorithm is
given $500$ impressions to learn from the same distribution of
impressions.

The minimum prices of each impression come from the range $[0.2R,R]$
for most of the simulations, but we also verified our broad findings
when the range of minimum prices is $[0.5R,R]$. Higher minimum prices
reflect lower levels of maximum possible sales.

For our set-based algorithms, we try out all powers of $2$ for the
value of $k$, that is, $1,2,4,8,16$ and $32$. For our threshold
algorithms, we try $0.5, 1.0,1.5,2.0,2.5$ and $3.0$ as our threshold
values.
This completes the description of the simulation set up, and we are
now ready to describe our findings.

\begin{figure*}
\begin{tabular}{cc}
\includegraphics[scale=0.2]{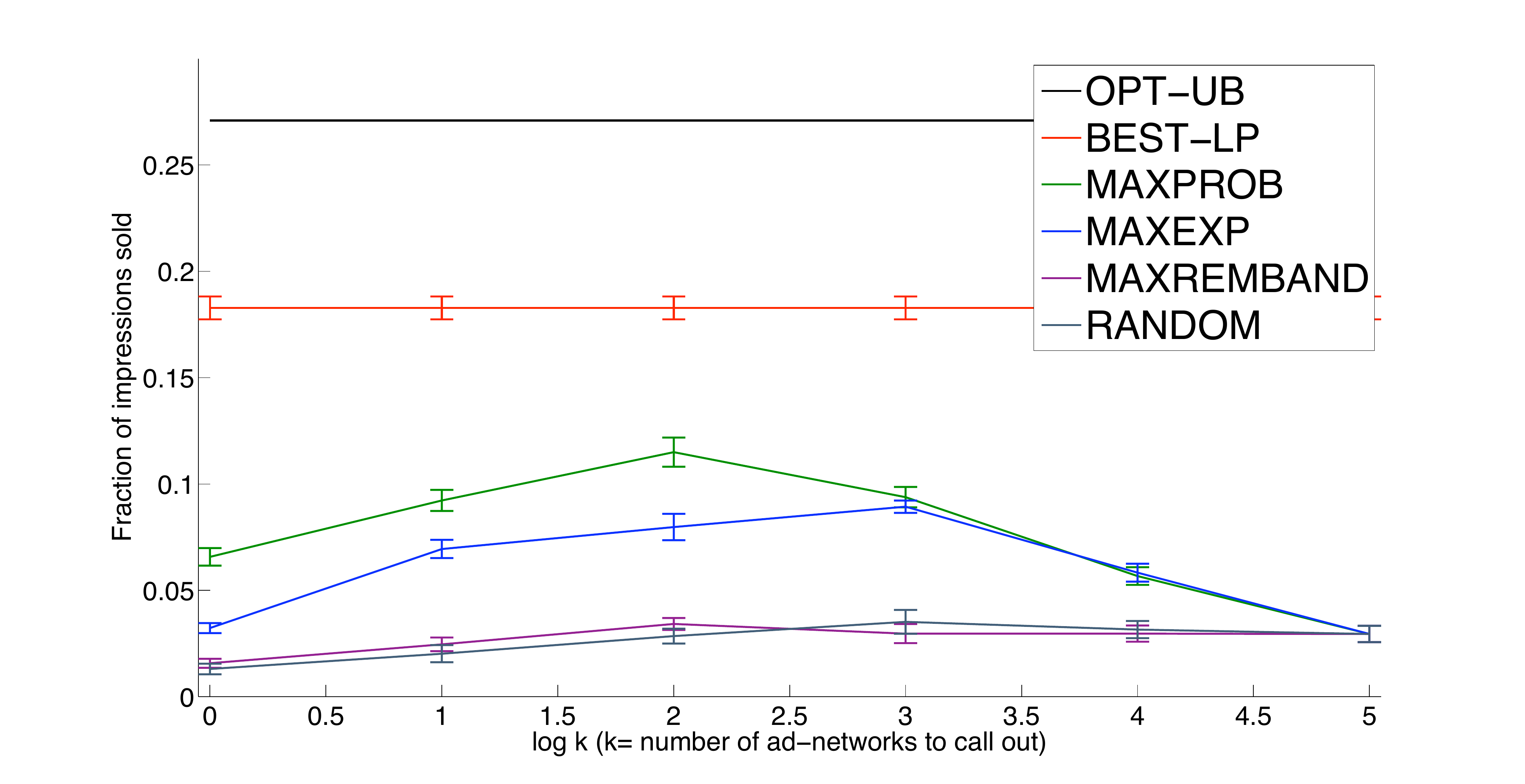}
&
\includegraphics[scale=0.2]{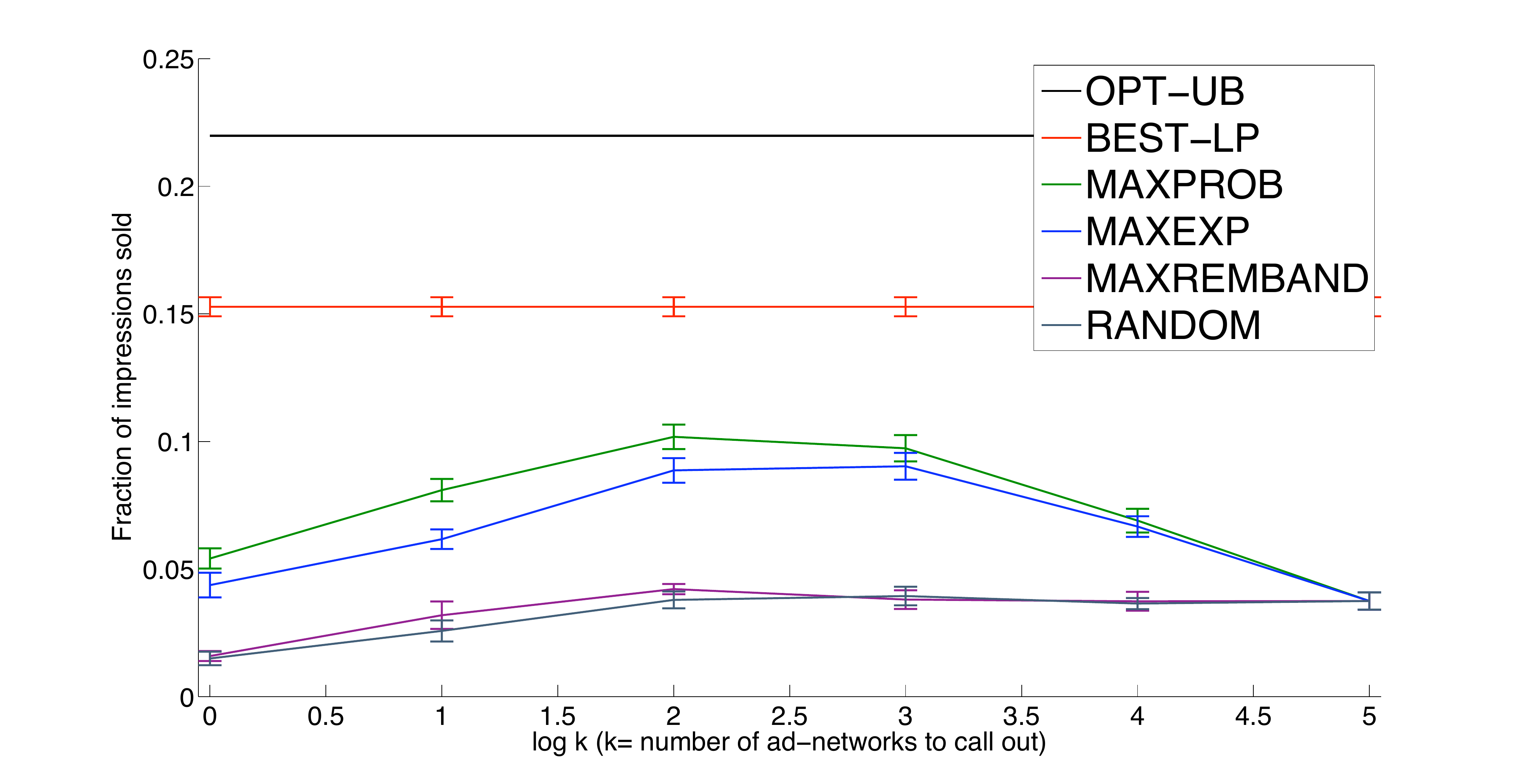}\\
(Gaussian) & (Pareto)
\end{tabular}
\caption{Set-based algorithms for bid distributions with
 higher minimum prices}\label{GP0010}
\end{figure*}

\subsection{Performance of various Strategies}
We tried our algorithms for 2000 impressions in each simulation. While
this stream may seem small, we checked the standard deviation of our
results when different streams drawn from the same distribution were
used, and found them to be sufficiently small. With $10$ different
such streams, the standard deviation was found to be about $2\%$ or
less for all the algorithms. The performance (along with error bars
indicating the deviation) of the various strategies are plotted in
Figures \ref{G0000}, \ref{P0000} and \ref{GP0010}. While the performance of
threshold algorithms are plotted against the threshold parameter,
that of the set based algorithms are plotted against $k$ on a
logarithmic scale, that is, $\log k$. BEST-LP indicates the
performance of our LP-based algorithm for optimal choice of
threshold, and is shown along with set based algorithms for
comparison. OPT-UB indicates the upper bound on an optimal policy that
is given by LP1 in the previous
section. Note that the real offline optimum can be
significantly smaller, but is hard to compute.
The following conclusions emerge:
\begin{figure*}
\begin{tabular}{cc}
\includegraphics[scale=0.2]{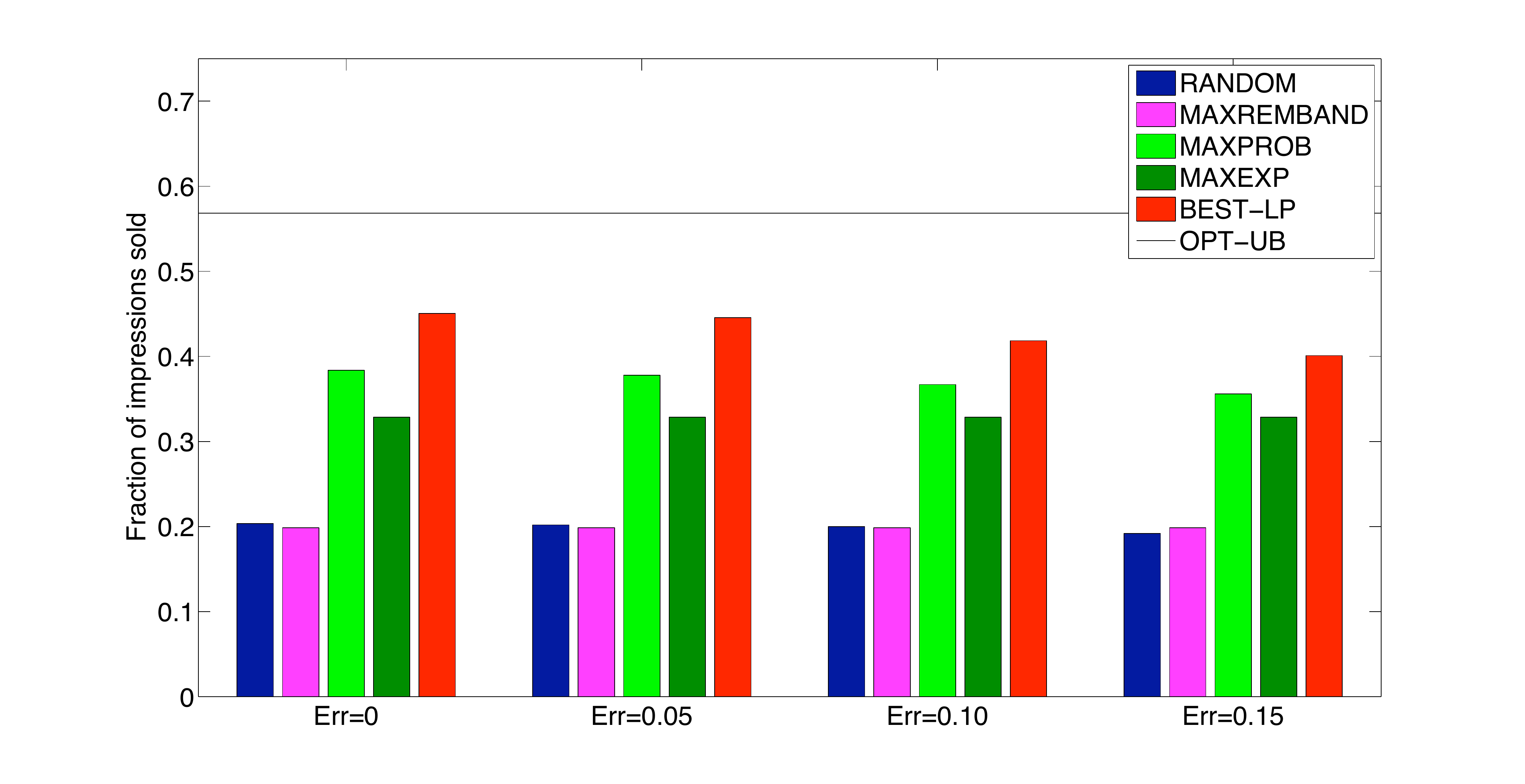}
&
\includegraphics[scale=0.2]{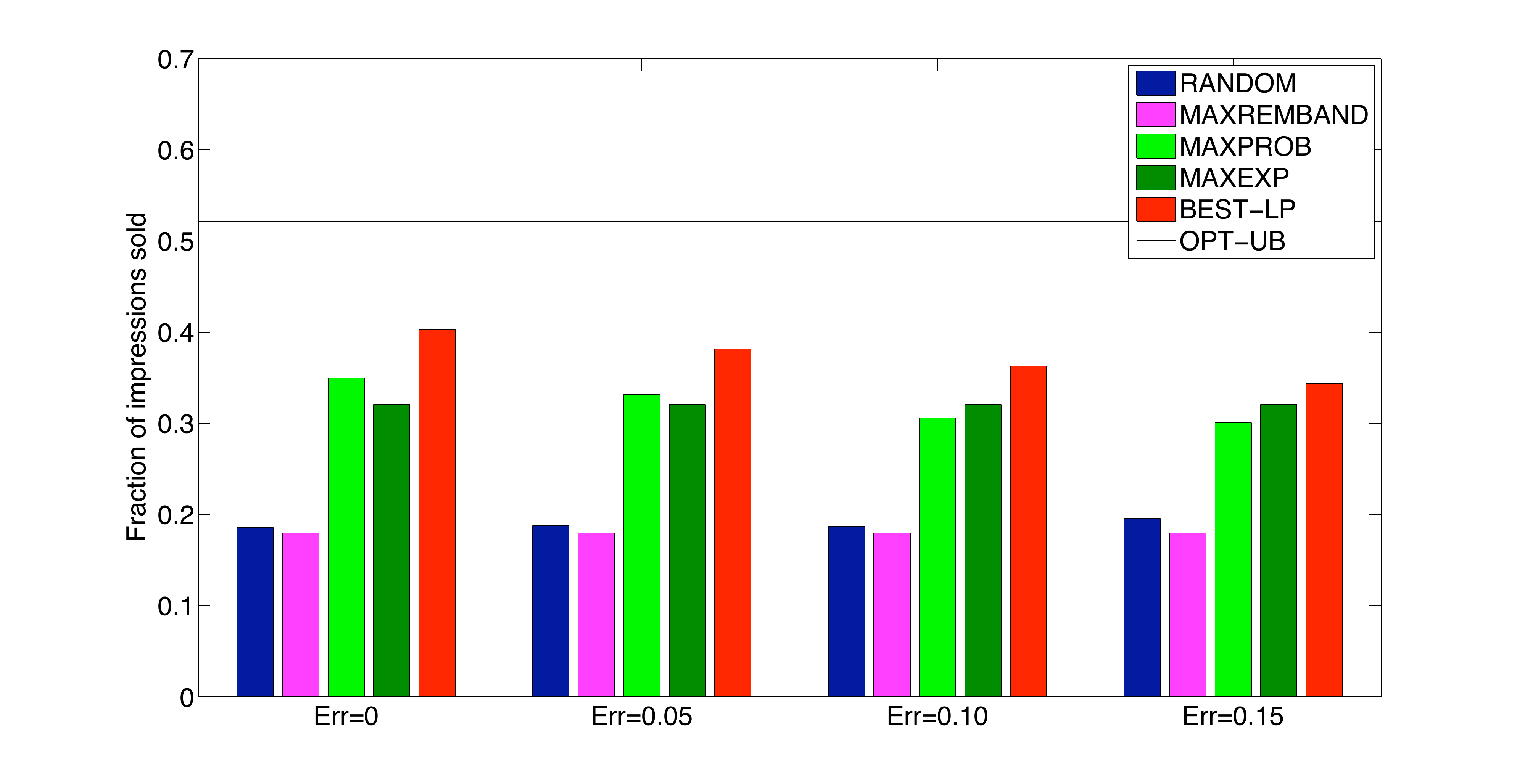}\\
(Gaussian) & (Pareto)
\end{tabular}
\caption{Set-based algorithms for different bid distributions with
 erroneous bid estimations}\label{error_GP}
\end{figure*}

\begin{enumerate}
\item The LP-based algorithm surpasses the performance of the other algorithms
by a large margin, about $20\%$ more than the closest performer with
optimal parameter choice. This gap grows to $85\%$ when the reserve
prices are raised, and the performance of all the algorithms (and also
the optimum) falls. Thus, our LP-based algorithm is even more valuable
in a world where fewer items can be sold. Figure ~\ref{GP0010},
show the effect of raising the minimum prices for set
based algorithms.
\item The algorithms Random and MaxRemBand perform quite poorly
  compared to
  all other algorithms that use information about bid
  distributions. This shows that {\em bid estimation is useful.}
\item The performance of MaxExp was better than Random and MaxRemBand,
  but was worse than MaxProb. Thus for the distribution used in the
  simulation, information about expectation was useful, but less useful
  than the survival probability of the bids.
\item The other threshold based algorithms perform quite poorly. In
  the rest of the experimental evaluation, we will ignore these
  algorithms.
\end{enumerate}

\subsection{Robustness and Error estimation}

Estimating survival probabilities can be a tough problem, and can be
loosely compared to the problem of estimating click-through-rates. It
is thus very likely that the estimates will be inaccurate, and it is
very important to understand how useful such erroneous estimates
are. We add noises to the survival probabilities that are normally
distributed (round to $0$ or $1$ if the resulting value is below $0$
or above $1$, respectively), and have standard deviations ranging from
$0.05$ to $0.15$. We find that the algorithms using this information
lose some performance, but still perform consistently better than the
algorithms without any information about bid distributions. We show
the plots for both distributions, in Figure~\ref{error_GP}. For
comparison, we choose to represent an algorithm by its peak
performance, that is, its performance for optimal choice of
parameters.


\begin{figure}[h]
\begin{minipage}{\textwidth}
\centerline{\includegraphics[scale=0.2]{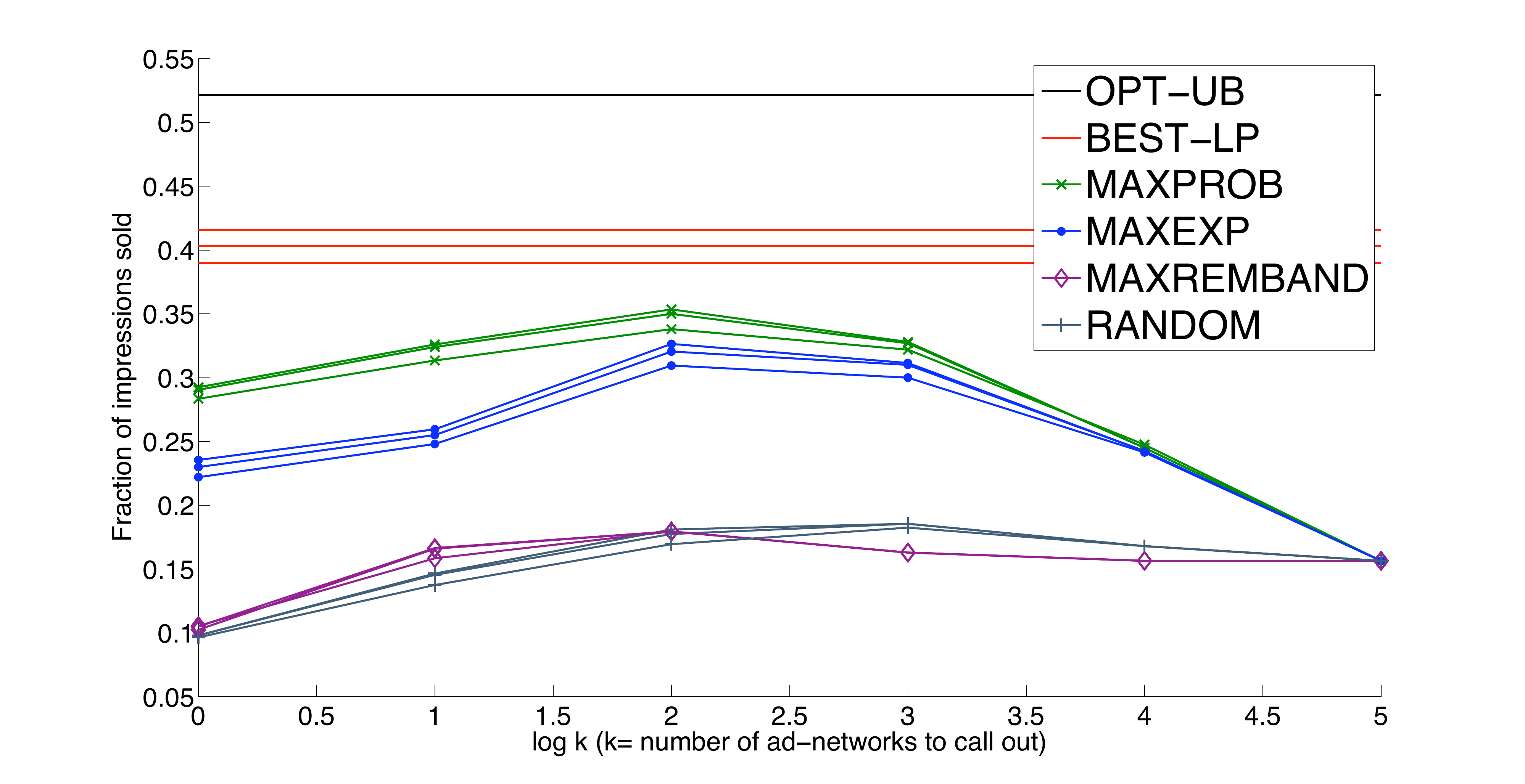}}
\caption{Set based algorithms for Pareto bid distributions with
 burst sizes $2$, $5$ and $15$}\label{buffer_P_k}
\end{minipage}
\end{figure}

\subsection{Sensitivity analysis}
\label{sensitivity}

\begin{figure}[h]
\begin{minipage}{\textwidth}
\centerline{\includegraphics[scale=0.2]{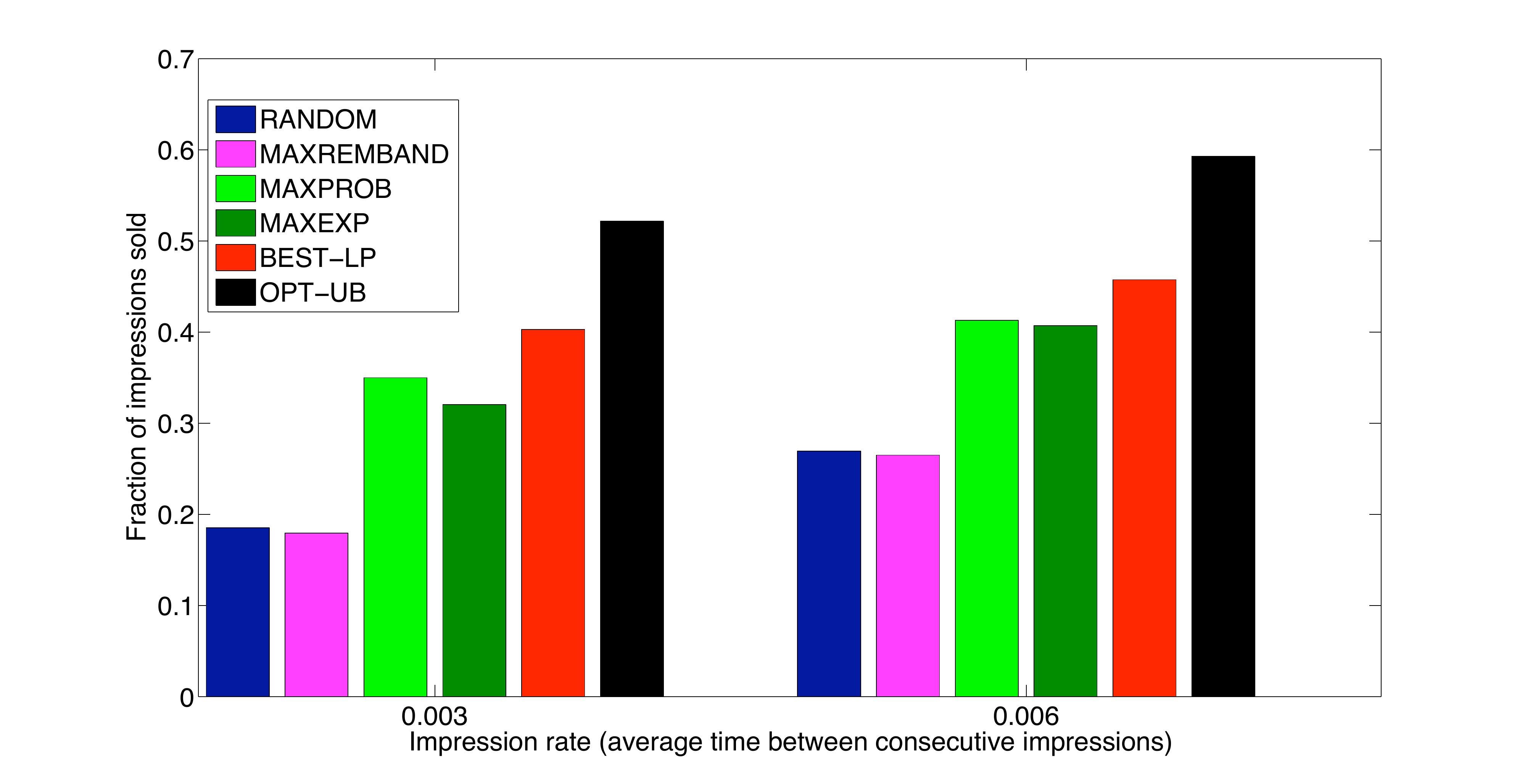}}
\caption{Peak performances of algorithms for different impression
  arrival rates for Pareto bid distributions}\label{slotrate_P}
\end{minipage}
\end{figure}

For all the plots described above, the size of the token buckets
were fixed at $5$ and the arrival rate was fixed to $0.003$.  We
tried out bucket sizes of $2$, $5$, $15$, and $45$.  As expected,
the performance of each algorithm is greater when bucket size is
larger; but there was no substantial differences in the conclusions
we observed. In fact, the performance of the algorithms were within
a standard deviation of a different bucket size (we omit showing
them to avoid clutter) There was no difference in performance
between bucket sizes $15$ and $45$. We superimpose the performance
plots of bucket sizes $2,5,15$ and show this plot for Pareto
distributions in Figure~\ref{buffer_P_k}. We also varied impression
arrival rate to note its effect on our algorithms. We found, as
expected, that performance improves if arrival rate slows down, but
the relative performances of the algorithms remain unchanged.
Figure~\ref{slotrate_P} shows this for the peak performances of the
algorithms, for Pareto bid distributions.

The
plots for Gaussian distributions and threshold algorithms are similar in
both cases.

\section{Conclusion \& Future Work}
\label{sec:conclusion}
We initiate a formal study of bandwidth and resource constraints faced
by ad exchanges and ad networks as they move into the arena of real
time bid solicitation. Our conceptual framework
of learning bid
distribution followed by online decision making
should be useful for
further study of real time bidding systems.
Our online algorithms are
fast and can be implemented to run in real time within this conceptual
framework. The technical framework of our algorithm, with a short
learning phase followed by a long online phase, can generally be
applied to any problem that involves maximizing linear objectives, or
objectives that can be approximated by a concave or linear objective,
with linear constraints.

\bibliographystyle{abbrv}
\bibliography{cop}

\begin{thebibliography}{10}

\bibitem{Azar}
Y.~Azar, B.~E. Birnbaum, A.~R. Karlin, and C.~T. Nguyen.
\newblock On revenue maximization in second-price ad auctions.
\newblock {\em Proc. of ESA}, pages 155--166, 2009.

\bibitem{BIKR07}
M.~Babaioff, N.~Immorlica, D.~Kempe, and R.~Kleinberg.
\newblock Knapsack secretary problem with applications.
\newblock {\em Proceedings of Approx}, pages 16--28, 2007.

\bibitem{barlow}
R.~E. Barlow, A.~W. Marshall, and F.~Proschan.
\newblock Properties of probability distributions with monotone hazard rate.
\newblock {\em The Annals of Mathematical Statistics}, 34(2):375--389, 1963.

\bibitem{sayan}
S.~Bhattacharya, G.~Goel, S.~Gollapudi, and K.~Munagala.
\newblock Budget constrained auctions with heterogeneous items.
\newblock {\em STOC, also at http://arxiv.org/abs/0907.4166}, 2010.

\bibitem{BJN}
N.~Buchbinder, K.~Jain, and J.~Naor.
\newblock Online primal-dual algorithms for maximizing ad-auctions revenue.
\newblock {\em Proc. of ESA}, pages 253--264, 2007.

\bibitem{Calinescu}
G.~Calinescu, C.~Chekuri, M.~P\'{a}l, and J.~Vondr\'{a}k.
\newblock Maximizing a submodular set function subject to a matroid constraint
  (extended abstract).
\newblock {\em IPCO}, pages 182--196, 2007.

\bibitem{CG08}
D.~Chakrabarty and G.~Goel.
\newblock On the approximability of budgeted allocations and improved lower
  bounds for submodular welfare maximization and gap.
\newblock {\em Proceedings of FOCS}, pages 687--696, 2008.

\bibitem{chawla}
S.~Chawla, J.~Hartline, D.~Malec, and B.~Sivan.
\newblock Sequential posted pricing and multi-parameter mechanism design.
\newblock {\em STOC, also at http://arxiv.org/abs/0907.2435}, 2010.

\bibitem{DH09}
N.~Devanur and T.~Hayes.
\newblock The adwords problem: Online keyword matching with budgeted bidders
  under random permutations.
\newblock In {\em Proc. 10th Annual ACM Conference on Electronic Commerge
  (EC)}, 2009.

\bibitem{Sergei2}
A.~Ghosh, P.~McAfee, K.~Papineni, and S.~Vassilvitskii.
\newblock Bidding for representative allocations for display advertising.
\newblock In {\em Proc. of 5th Workshop on Internet and Economics(WINE)}, 2009.

\bibitem{Sergei}
A.~Ghosh, B.~I.~P. Rubinstein, S.~Vassilvitskii, and M.~Zinkevich.
\newblock Adaptive bidding for display advertising.
\newblock In {\em WWW}, pages 251--260, 2009.

\bibitem{GM06}
S.~Guha and A.~McGregor.
\newblock Approximate quantiles and the order of the stream.
\newblock {\em Proc. of PODS}, pages 273--279, 2006.

\bibitem{adaptive}
S.~Guha and K.~Munagala.
\newblock Adaptive uncertainty resolution in bayesian combinatorial
  optimization problems.
\newblock {\em ACM Transactions of Algorithms}, (to appear).

\bibitem{KRT98}
J.~Kleinberg, Y.~Rabani, and E.~Tardos.
\newblock Allocating bandwidth for bursty connections.
\newblock {\em SIAM J. Of Computing}, 30(1):171--217, 2001.

\bibitem{Kleinberg05}
R.~Kleinberg.
\newblock A multiple-choice secretary problem with applications to online
  auctions.
\newblock {\em Proceedings of SODA}, pages 630--631, 2005.

\bibitem{KSU08}
R.~Kleinberg, A.~Slivkins, and E.~Upfal.
\newblock Multi-armed bandits in metric spaces.
\newblock In {\em STOC}, pages 681--690, 2008.

\bibitem{Kulik}
A.~Kulik, H.~Shachnai, and T.~Tamir.
\newblock Maximizing submodular set functions subject to multiple linear
  constraints.
\newblock {\em SODA}, pages 545--554, 2009.

\bibitem{MV95}
A.~Marchetti-Spaccamela and C.~Vercellis.
\newblock Stochastic on-line knapsack problems.
\newblock {\em Mathematical Programming}, 68:73--104, 1995.

\bibitem{MSVV05}
A.~Mehta, A.~Saberi, U.~Vazirani, and V.~Vazirani.
\newblock Adwords and generalized on-line matching.
\newblock In {\em FOCS}, pages 264--273, 2005.

\bibitem{MuthuADX}
S.~Muthukrishnan.
\newblock Adexchange: Research issues.
\newblock In {\em Proc. of 5th Workshop on Internet and Economics(WINE)}, 2009.

\bibitem{Myerson}
R.~B. Myerson.
\newblock Optimal auction design.
\newblock {\em Math. of OR}, 6(1):58–73, 1981.

\bibitem{rtb}
M.~Nolet.
\newblock http://www.mikeonads.com/2009/08/30/rtb-part-i-what-is-it/.

\bibitem{PandeyOlston}
S.~Pandey and C.~Olston.
\newblock Handling advertisements of unknown quality in search advertising.
\newblock {\em Proceedings of NIPS}, pages 1065--1072, 2006.

\bibitem{DCLK}
P.~release.
\newblock http://www.doubleclick.com/insight/
  blog/archives/doubleclick-advertising-exchange/announcing-the-new-doubleclic%
k-ad-exchange.html.

\bibitem{rockafellar}
R.~T. Rockafellar.
\newblock {\em Convex Analysis}.
\newblock Princeton University Press, 1970.

\bibitem{Tanenbaum}
A.~S. Tanenbaum.
\newblock {\em Computer Networks, 3rd Edition}.
\newblock Prentice-Hall, 1996.

\bibitem{VVS10}
E.~Vee, S.~Vassilvitskii, and J.~Shanmugasundaram.
\newblock Optimal online assignment with forecasts.
\newblock In {\em EC}, 2010.

\end{thebibliography}
\appendix

\section{The Adversarial Model}
\label{adversarial}
We show that no online algorithm, deterministic or randomized, can
give any approximation to the offline optimum, even when the instance
has only one ad network, and the bids are $0$ or $1$.

\begin{theorem}
For any given $0<\alpha\leq 1$, there exists no randomized algorithm
which is $\alpha$-approximate, if the length of the sequence is
arbitrarily large.
\end{theorem}
\begin{proof}
Maximizing the sales objective in the presence of only one ad network
subsumes the online knapsack problem (the value of an item is the
probability that the bid exceeds minimum price) with unit size of
items, for which no
algorithm can give better than $1/\Omega(\log m)$ approximation
\cite{MV95}. If $m$ is allowed to be large, then no algorithm giving
approximation independent of $m$ can exist.
\end{proof}

One key feature of the adversarially constructed sequences in the
theorem above is that $\opt$ can be arbitrarily small compared to the
length of the sequence. This is unlikely to happen; moreover, if it
were to happen, then there would not be much benefit out of
constructing the real time bidding system. Thus we make the assumption
that $\opt$ is reasonably large. But even with that assumption, we
don't fare particularly well. Following is a simple 
observation about sales.

\begin{lemma}\label{lem-basic-sales}
For a given impression $j$, suppose ad network $i$, with survival
probability $p_{ij}$ is called out with probability $x_{ij}$. Then the
probability of selling the impression is $1 - \prod_{i=1}^n (1 - p_{ij}
x_{ij})$. If $\sum_{i=1}^n p_{ij} x_{ij}=c$, then the probability of selling the
impression is at most $c$ and at least $1-\frac{1}{e^c}$.
\end{lemma}
\begin{proof}
The probability of selling the impression is $1 - \prod_{i=1}^n (1-p_{ij}
x_{ij})$. It is upper bounded by $\sum_i p_{ij}x_{ij} =c$. The
probability is lower bounded by $1 - \prod_{i=1}^n (1-c/n)^n$ where
$n$ is the number of ad-networks, and this is lower bounded by $1 -
\frac{1}{e^c}$.
\end{proof}

\begin{theorem}
Suppose it is promised that $\opt \geq \delta m$, where $\delta>0$ is
known to the algorithm. Then, no algorithm can be better than
$1/\Omega(\log (1/\delta))$-approximate for the sales or total value
objective. Moreover, a simple randomized algorithm is $1/O(\log
(n/\delta))$-approximate for the sales objective.
\end{theorem}
\begin{proof}
The hardness instances for online knapsack problems also imply that if
the value of all items are guaranteed to lie between $L$ and $U$,
$L< U$,  then no algorithm can give better than $1/\Omega(\log (U/L))$
approximation \cite{MV95}, if $m$ is allowed to be arbitrarily
large. The promise $\opt\geq \delta m$ allows us to construct any
instance where values of items range from $\delta$ to $1$, thus
yielding the lower bound.
For the upper bound on sales, we use the following simple algorithm:
Let us assume, wlog, that $\delta$ is a negative power of $2$. Choose a
random cut-off $t$ from the set $H=\{\delta/2, \delta, 2\delta, 4\delta\ldots
1\}$. Now send each impression to any $2/t$ ad networks who has
survival probability between $t$ and $2t$ and has bandwidth
remaining. If there are less than $2/t$ such qualifying ad networks
for an impression, then send it to all qualifying ad networks. This
algorithm gives a $1/O(\log (n/\delta))$ approximation; the proof is
standard.
\end{proof}

\end{document}